\documentclass[letterpaper]{JHEP3}
\usepackage{array}
\usepackage{nicefrac}
\usepackage{amsmath,epsfig,axodraw}
\usepackage{amssymb,amsfonts,mathrsfs}
\usepackage{color}
\usepackage{skak}

\usepackage{graphics}
\usepackage{epsfig}
\usepackage{amsfonts}

\newcommand{\Tr}{\mbox{Tr}}

\newcommand{\Q}{{\cal Q}}
\global \long \def \NN{ \mathcal{N}}
\global \long \def \II{ \mathcal{I}}
\def\a{\alpha}
\def\b{\beta}
\def\e{\epsilon}

\def\h{\eta}

\def\half{{\frac12}}

\def\IC{\relax\hbox{$\inbar\kern-.3em{\rm C}$}}

\def\IC{{\bf C}}

\def\bea{\begin{eqnarray}}
\def\eea{\end{eqnarray}}
\def\be{\begin{equation}}
\def\ee{\end{equation}}
\def\ba{\begin{align}}
\def\ea{\end{align}}
\def\bse{\begin{subequations}}
\def\ese{\end{subequations}}
\def\1F1{{}_1\!F_1}
\def\2F0{{}_2\!F_0}

\def\a{\alpha}

\def\h3{$\textrm{H}_3^+$}

\def\IC{{\mathbb C}}

\def\Tr{{\rm Tr}}

\def\lbldef#1#2{\expandafter\gdef\csname #1\endcsname {#2}}

\def\href#1#2{#2}

\newcommand{\beq}{\begin{equation}}
\newcommand{\eeq}{\end{equation}}
\newcommand{\ber}{\begin{eqnarray}}
\newcommand{\eer}{\end{eqnarray}}

\def\be{\begin{eqnarray}}
\def\ee{\end{eqnarray}}

\providecommand{\tabularnewline}{\\}

\def\({\left(}
\def\){\right)}
\def\[{\left[}
\def\]{\right]}
\def\<{\langle}
\def\>{\rangle}

\def\BB{{\cal B}}
\def\QQ{{\cal Q}}
\def\CC{{\cal C}}
\def\DD{{\cal D}}
\def\AA{{\cal A}}
\def\II{{\cal I}}

\title{On the Superconformal Index of ${\cal N}=1$ IR Fixed Points \\ \centerline{\Large A Holographic Check} }
\preprint{YITP-SB-10-38}
\author{Abhijit Gadde\footnote{abhijit@insti.physics.sunysb.edu},
 Leonardo Rastelli\footnote{leonardo.rastelli@stonybrook.edu},
Shlomo S. Razamat\footnote{razamat@max2.physics.sunysb.edu}, and
Wenbin Yan\footnote{wyan@insti.physics.sunysb.edu}
\\
\\
\it C.N. Yang Institute for Theoretical Physics,\\
\it Stony Brook University, \\
\it Stony Brook, NY 11794-3840, USA}

\bigskip

\abstract{We evaluate the superconformal index of the $Y^{p,q}$ quiver gauge theories using R\"omeslberger's prescription.
For the conifold quiver $Y^{1,0}$ we find exact agreement at large $N$ with a previous calculation in the dual  $AdS_5 \times T^{1,1}$
supergravity.
}

\begin{document}

\section{Introduction}

The superconformal index~\cite{Kinney:2005ej} encodes
``robust''  information about the protected states of a superconformal field theory (SCFT).
 It is a weighted sum over the states of the theory, which by construction evaluates to zero on a generic (long) multiplet. It follows that
 the index is invariant under exactly marginal deformations, since it is not affected by the recombinations of short multiplets into long ones (or viceversa) that may occur as parameters are varied.
For SCFTs admitting a weakly-coupled limit, the index can then be evaluated in  free-field theory by a straightforward counting procedure. It takes the form
of a matrix integral.

This is much less trivial than it sounds. For $4d$ SCFTs with ${\cal N}=4$ and ${\cal N}=2$ supersymmetry,
the  evaluation of the index can often be carried out in different weakly-coupled frames related
by $S$-duality, leading to different--looking, but equivalent integral representations of the same index \cite{Gadde:2009kb, Gadde:2010te}. 
These different representations are related
by   identities between elliptic hypergeometric integrals, an active  subject of  mathematical research. 
 $S$-duality of the index can also be phrased
 as associativity of the operator algebra of a $2d$ topological QFT \cite{Gadde:2009kb}. This line of thought
  gives also  a way to evaluate (in principle) the index of some  SCFTs with no Lagrangian description, by relating them to weakly-coupled theories:
a concrete example is the $E_6$ SCFT,
whose index was found in closed form in \cite{Gadde:2010te} by demanding consistency with Argyres-Seiberg duality.

 On the other hand, some of the most important examples of  interacting $4d$ SCFTs
do not have a (known) weakly-coupled description in any duality frame. A large class
are the ${\cal N}=1$ SCFTs that arise as IR fixed points of renormalization group flows, whose UV starting points
are weakly-coupled theories. A prescription to evaluate the index of such SCFTs was
formulated by R\"omelsberger \cite{Romelsberger:2005eg, Romelsberger:2007ec}.
This prescription has so far been checked indirectly, by showing in several examples that it gives
the same result for different RG flows that end in the same IR fixed point ({\it i.e.} the UV theories are Seiberg dual).
This was originally observed
 by R\"{o}melsberger, who performed a few  perturbative checks in a chemical potential expansion \cite{Romelsberger:2005eg, Romelsberger:2007ec}.
 Invariance of the ${\cal N}=1$ index under Seiberg duality was systematically demonstrated
 by Dolan and Osborn~\cite{Dolan:2008qi}, in a remarkable paper that first applied the elliptic hypergeometric machinery
 to the evaluation of the superconformal index. These  results were extended and generalized
in~\cite{Spiridonov:2008zr,Spiridonov:2009za,Spiridonov:2010hh,Vartanov:2010xj}.

In this  note we apply  R\"{o}melsberger's prescription to a class
  of     ${\cal N}=1$ SCFTs that admit AdS duals.
The canonical example
is the conifold theory of Klebanov and Witten~\cite{Klebanov:1998hh}. There are infinitely many generalizations: the  families of toric quivers
 $Y^{p,q}$~\cite{Benvenuti:2004dy} and  $L^{p,q,r}$~\cite{Cvetic:2005ft}. We focus on  $Y^{p,q}$.
In all these examples
there is in principle an independent way to determine  the  index (at large $N$)
from the dual supergravity.
We will explicitly show agreement between the gravity calculation of Nakayama \cite{Nakayama:2006ur} and
our field theory calculation  for the case of the conifold quiver $Y^{1,0}$. According to taste, this can be viewed either as
a check of   R\"{o}melsberger's prescription, or as yet another check of AdS/CFT. The upshot is
 a sharper bulk/boundary dictionary.

To make the paper self-contained, we review in section 2   the ${\cal N}=1$ superconformal index and
R\"{o}melsberger's prescription, making some comments on its rationale.
The  idea (implicit in the discussion of \cite{Romelsberger:2005eg, Romelsberger:2007ec}) is to re-interpret the superconformal index of the IR theory as
the Witten index of the non-conformal theory on $S^3 \times \mathbb{R}$ describing the whole RG flow.
In section 3 we present a simple universal relation  between the indices
 of a UV ${\cal N}=2$ and an IR ${\cal N}=1$ SCFTs connected by the RG flow triggered by a mass term for the adjoint chiral superfield.
In section~\ref{Ypqsec} we  review basic facts
about the $Y^{p,q}$ family of toric quivers (the conifold being a special case $Y^{1,0}$). From the quiver diagrams, it is immediate
to write integral expressions for the  superconformal index,  at finite $N$. We show that the indices of toric-dual theories are equal,
as expected.
In section \ref{largeNsec} we consider the large $N$ limit. We conjecture  a simple closed
form expression for the large $N$ index of the $Y^{p,q}$ quivers.
In section~\ref{gravitysec} we review the gravity computation of the index for the conifold~\cite{Nakayama:2006ur} and find exact agreement with the large $N$ limit of our field theory result.
An appendix collects useful material about   ${\cal N}=1$ superconformal representation theory
and the  index of the different short and semishort supermultiplets.

\section{Review of the ${\cal N}=1$ index}

The  index of a $4d$ superconformal field theory is defined
as the Witten index of the theory in radial quantization.  Let ${\cal Q}$ be  one of the  Poincar\'e supercharges, and $\Q^\dagger ={\cal  S}$
the conjugate conformal supercharge. Schematically, the index is defined as ~\cite{Kinney:2005ej, Romelsberger:2005eg, Romelsberger:2007ec}
 \be \label{basicdef} {\mathcal I} (\mu_i)={\rm
Tr}\, (-1)^{F}\,e^{-\beta\, \delta}\, e^{-\mu_i {\cal M}_i}\, ,
\ee
where the trace is over the Hilbert space  of the theory on $S^3$,
$\delta \equiv \frac{1}{2} \{\Q,\, \Q^{\dagger}\}$,
  ${\cal M}_i$ are $\Q$-closed
conserved charges and $\mu_i$ the associated chemical potentials. Since states with $\delta >0$
come in boson/fermion pairs, only the $\delta =0$ states contribute, and the index is independent of $\beta$.
There are  infinitely many states with $\delta =0$ --
this is true even  for a single  short irreducible representation of the superconformal algebra,
 because
some of the non-compact generators (some of the spacetime derivatives)
  have $\delta=0$. The introduction of the chemical
potentials $\mu_i$ serves both to regulate this divergence and to achieve a more
refined counting.

For  ${\cal N}=1$, the supercharges are $\{ {\cal Q}_\alpha \, , {\cal S}^\alpha \equiv {\cal Q}^{\dagger\, \alpha} \, , {\widetilde {\cal Q}_{\dot \alpha}} \, ,
\widetilde {\cal S}^{\dot \alpha} \equiv {\widetilde {\cal Q}^{\dagger \,\dot \alpha}} \}$,
where $\alpha = \pm$ and $\dot \alpha = \dot \pm$ are respectively $SU(2)_1$ and $SU(2)_2$ indices, with  $SU(2)_1 \times SU(2)_2 = Spin(4)$ the isometry group of the $S^3$.
The relevant anticommutators are
\bea
\{\Q_\alpha, \, {\cal \Q}^{\dagger\, \beta} \} & =& E+2M_\alpha^\beta+\frac{3}{2}r \\
\{\widetilde \Q_{\dot \alpha}\,, \, {\widetilde {\cal \Q}}^{\dagger \, \dot \beta} \} & =& E +2 \widetilde M_{\dot \alpha}^{\dot \beta}-\frac{3}{2}r \, ,
\eea
where $E$ is the conformal Hamiltonian, $M_{\alpha}^\beta$ and  $\widetilde M_{\dot \alpha}^{\dot \beta}$  the $SU(2)_1$ and $SU(2)_2$ generators, and
$r$ the generator of the $U(1)_r$ R-symmetry. In our conventions, the $\Q$s have $r=-1$ and $\widetilde Q$s have $r=+1$,
and of course the dagger operation  flips the sign of $r$.

One can define two inequivalent indices,
a ``left-handed'' index   $\II^{{\tt L}}(t,y)$  and a ``right-handed'' index  $\II^{{\tt R}}(t,y)$. For the left-handed index,
we pick say\footnote{Picking $\Q \equiv\Q_{+}$ would amount to the replacement $j_1 \leftrightarrow -j_1$, which is an equivalent choice because of $SU(2)_1$ symmetry. The same consideration applies to the right-handed index, which can be defined either choosing $\widetilde \Q_{\dot -}$ or $\widetilde \Q_{\dot +}$. } $\Q \equiv \Q_{-}$:
\be
\label{eq:defofindexL}
\II^{{\tt L}}(t,y) \equiv {\rm Tr} \, (-1)^F t^{2(E+j_1)}y^{2j_2} =   {\rm Tr} \, (-1)^F t^{3(2j_1-r)}y^{2j_2}\, ,\qquad
\delta=E-2j_1+\frac{3}{2}r \, ,
\ee
where $j_1$ and $j_2$ are the Cartan generators of $SU(2)_1$ and $SU(2)_2$.
The two ways of writing the exponent of $t$ are equivalent since they differ by a $\Q$-exact term. For the right-handed index,
we pick
say $\Q \equiv \widetilde \Q_{\dot -}$
\be
\label{eq:defofindexR}
\II^{{\tt R}}(t,y) \equiv {\rm Tr} \, (-1)^F t^{2(E+j_2)}y^{2j_1} =   {\rm Tr} \, (-1)^F t^{3(2j_2+r)}y^{2j_1}\, ,\qquad
\delta=E-2j_2-\frac{3}{2}r \,.
\ee
One may also introduce chemical potentials for additional global symmetries of the theory.

\subsection{Romelsberger's prescription}

The expression (\ref{basicdef}) makes sense  for a general supersymmetric QFT on $S^3 \times \mathbb{R}$.
In particular we can consider a theory that flows between two conformal fixed points in the  UV and in the IR. At a fixed point (and only at a fixed point), the theory on $S^3 \times \mathbb{R}$ is equivalent to a superconformal  theory on $\mathbb{R}^4$, and $Q^\dagger$
can be interpreted as a conformal supercharge on $\mathbb{R}^4$. By the usual formal arguments, the index
is invariant along the flow (it is independent of the dimensionless coupling $R M$, where $R$ is the $S^3$ radius
and $M$ the renormalization group scale).  For this procedure to make sense, clearly the $Q$-closed charges ${\cal M}_i$ must be
well-defined (in particular non-anomalous) all along the RG flow.
 If the UV fixed point is a free theory, we can compute its index by a matrix
integral that counts the gauge-invariant words  with $\delta_{UV}=0$. We can then re-intepret the result as the superconformal index
of the IR fixed point, which would be difficult to evaluate directly.  This leads to the following prescription \cite{Romelsberger:2007ec,Romelsberger:2005eg}
\begin{enumerate}
\item Consider the UV starting point. Write down the ``letters'' contributing to the index of the free theory, {\it i.e.}
the letters with $\delta_{UV}=0$.
\item Assign to the letters the quantum numbers corresponding to the anomaly-free symmetries of the interacting
theory. In the presence of $U(1)$ global symmetries,  follow  the usual $a$-maximization procedure~\cite{Intriligator:2003jj}
to single-out the anomaly-free $R$-symmetry that in the IR becomes  the $U(1)_r$ of the superconformal algebra.
\item Compute the index in terms of  the matrix integral which enumerates  gauge-invariant words.
\end{enumerate}
The considerations leading to this recipe are somewhat formal. One direction in which they could be made more precise
is to discuss ultraviolet regularization and renormalization.
It is not difficult to find a perturbative regulator that preserves one complex $\Q$, and in fact two of them,
 either the two left-handed charges $\Q_\alpha$, or the two right-handed charges $\widetilde Q_{\dot \alpha}$.
 To preserve say the left-handed charges, we can Kaluza-Klein expand
  the fields on the $S^3$, and truncate the theory by keeping all the modes whose   right-handed
 spin $J_2 \leq J_2^{max}$. This truncation is a UV regulator since the left-handed modes
 will also be cut-off\footnote{This is clear from the structure of harmonics on $S^3$. Scalar harmonics
 have $SU(2)_1 \times SU(2)_2$ quantum numbers $(J, J)$, spinor harmonics $(J-1/2, J)$ and $(J, J-1/2)$ and so on.},
 and has the virtue of preserving the left-handed supersymmetry,
 since the action of $\Q_\alpha$ commutes with the cut-off. A similar regulator (but performed
 symmetrically on the left-handed and right-handed spins, which in general breaks susy) has
been discussed at length in~\cite{Aharony:2005bq,Aharony:2006rf,Mussel:2009uw, Ishiki:2006rt}.
This style of regularization is only  perturbative because it breaks the
 gauge symmetry, which can however  be restored order by order in perturbation theory by adding  counterterms~\cite{Aharony:2005bq,Aharony:2006rf,Mussel:2009uw, Ishiki:2006rt}.  We see no obstacle in choosing the counterterms so that they preserve
 one copy of the susy algebra.

We are not aware of a fully non-perturbative
regulator that preserves supersymmetry on
 $S^3 \times \mathbb{R}$ -- finding such a regulator would be very interesting in its own right.
In any case  ultraviolet
issues  are not expected to affect the play an important role for the index, much as
 they don't  for the usual Witten index on the torus \cite{Witten:1982df}.

\subsection{Computing the index}

The   ``letters'' of an $N=1$ chiral multiplet
are enumerated
in table \ref{tab1}. We assume that in the IR the $U(1)_r$ charge
of the lowest component of the multiplet $\phi$ is some arbitrary $r_{IR}=r$
(determined in a concrete theory by anomaly cancellation and in subtle cases $a$-maximization).
According to the prescription we have just reviewed,
the index receives contributions from the letters with $\delta_{UV}=0$,
and each letter contributes as $(-1)^F t^{3(2j_1-r_{IR})}y^{2 j_2}$ to the left-handed
index and as  $(-1)^F t^{3(2j_2+ r_{IR})}y^{2 j_1}$  to the right-handed index.
\begin{table}[htbp]
\begin{centering}
\begin{tabular}{|c||c|c|c|c|c||c|c||c|c|}
\hline Letters & $E_{UV}$ & $j_{1}$ & $j_{2}$ & $r_{UV}$ & $r_{IR}$
& $\delta_{UV}^{{\tt L}}$ & ${\cal I}^{\tt L}$ & $\delta_{UV}^{{\tt
R}}$ & ${ \cal I}^{\tt R}$ \tabularnewline \hline \hline $\phi$ &
$1$ & $0$ & $0$ & $\frac{2}{3}$ & $r$ & $2$ & $-$ & $0$ &
$t^{3r}$\tabularnewline \hline $\psi$ & $\frac{3}{2}$ &
$\pm\frac{1}{2}$ & $0$ & $-\frac{1}{3}$ & $r-1$ & $0^+,2^-$ &
$-t^{3(2-r)}$ & $2$ & $-$ \tabularnewline \hline $\partial \psi$ &
$\frac{5}{2}$ & $0$ & $\pm\frac{1}{2}$ & $-\frac{1}{3}$ & $r-1$ &
$2$ & $-$
 & $4^+,2^-$ & $-$ \tabularnewline \hline $\square\phi$ & $3$ & $0$ & $0$ &
$\frac{2}{3}$ & $r$ & $4$ & $-$ & $2$ & $-$ \tabularnewline \hline
\hline $\bar{\phi}$ & $1$ & $0$ & $0$ & $-\frac{2}{3}$ & $-r$ & $0$
& $t^{3r}$ & $2$ & $-$\tabularnewline \hline $\bar{\psi}$ &
$\frac{3}{2}$ & $0$ & $\pm\frac{1}{2}$ & $\frac{1}{3}$ & $-r+1$ &
$2$ & $-$ & $2^+,0^-$ & $-t^{3(2-r)}$\tabularnewline \hline
$\partial\bar{\psi}$ & $\frac{5}{2}$ & $\pm\frac{1}{2}$ & $0$ &
$\frac{1}{3}$ & $-r+1$ & $2^+,4^-$ & $-$ & $2$ & $-$ \tabularnewline
\hline $\square\bar{\phi}$ & $3$ & $0$ & $0$ & $-\frac{2}{3}$ & $-r$
& $2$ & $-$
 & $4$ & $-$\tabularnewline \hline\hline
$\partial_{\pm\pm}$ & $1$ & $\pm\frac{1}{2}$ & $\pm\frac{1}{2}$ &
$0$ & $0$ & $0^{\pm+},2^{\pm-}$ & $t^3y^{\pm1}$ &
$0^{+\pm},2^{-\pm}$ & $t^3y^{\pm1}$\tabularnewline \hline
\end{tabular}
\par\end{centering}
\caption{\label{tab1}The ``letters'' of an ${\cal N}=1$ chiral
multiplet and their contributions to the index. Here $\delta^{\tt L}=E-2j_{1}+\frac{3}{2}r_{UV}$ and
$\delta^{\tt R}_{UV}=E-2j_{2}-\frac{3}{2}r_{UV}$.  A priori
we have to take into account the free equations of motion $\partial \psi =0$ and $\Box \phi =0$,
which imply constraints on the possible words,
but we see that in this case equations of motions have $\delta_{UV} \neq 0$ so they do not change the index. Finally there
are two spacetime derivatives contributing to the index, and their multiple
action on the fields is responsible for the denominator of the index, $1/(1-t^3 y^{\pm1}) = \sum_{n=0}^\infty (t^3 y^{\pm 1})^n$.
}
\end{table}
To keep track of the gauge and flavor quantum numbers,
we introduce  characters. We assume that the chiral multiplet
transforms in the representation   ${\cal R}$ of the gauge $\times$ flavor group,
and denote by $\chi_{\cal R}(U, V)$,  $\chi_{\bar {\cal R}}(U, V)$ the characters of  ${\cal R}$ and
and of the conjugate representation $\bar {\cal R}$, with $U$ and $V$  gauge and flavor group matrices respectively.
All in all, the single-letter left- and right-handed indices for a chiral multiplet are \cite{Dolan:2008qi}
\bea
 i_{\chi(r)}^{\tt L}(t,y,U, V) &  =&
\frac{t^{3r}\,\chi_{\bar{\mathcal R}}(U, V)-t^{3(2-r)}\,\chi_{\mathcal
R}(U, V)}{(1-t^{3}y)(1-t^{3}y^{-1})} \\  
\label{iC1}
  i_{\chi(r)}^{\tt R}(t,y,U, V) &  = &
\frac{t^{3r}\,\chi_{{\mathcal R}}(U, V)-t^{3(2-r)}\,\chi_{\bar {\mathcal
R}}(U,V)}{(1-t^{3}y)(1-t^{3}y^{-1})}\,.   \ee
The denominators encode the action of the two spacetime derivatives with $\delta = 0$.
Note that the left-handed and right-handed indices differ by conjugation of the gauge and flavor quantum numbers.
As a basic consistency check \cite{Romelsberger:2007ec}, consider a single free massive chiral multiplet (no gauge or flavor indices).
In the UV, we neglect the mass deformation and as always $r_{UV} = \frac{2}{3}$. In the IR, the quadratic superpotential
implies $r_{IR}=1$, and one finds  $i_{r=1}^{\tt L} =  i_{r=1}^{\tt R} \equiv 0$. As expected, a massive superfield decouples and
does not contribute to the IR index.

Finding the contribution to the index of an ${\cal N}=1$ vector multiplet  is even easier,
since the $R$-charge of a vector superfield $W_\alpha$ is fixed at the canonical value $+1$
all along the flow.
For  both left- and the right-handed index,
the single-letter index of a vector multiplet is
~\cite{Kinney:2005ej}
\be \label{iV1}
i_{V}(t,y,U)  =
\frac{2t^{6}-t^{3}(y+\frac{1}{y})}{(1-t^{3}y)(1-t^{3}y^{-1})}\,\chi_{adj}(U) \, . \ee

Armed with the single-letter indices, the full index is obtained by enumerating all the
words and then projecting onto gauge-singlets
 by integrating over the Haar measure of the gauge group.
Schematically,
\begin{equation}\label{integral}
{\cal I}(t, y, V)=\int[dU] \, \prod_k  \,  \, {\rm PE}[i_k (t, y, U, V)] \,,
\end{equation}
where $k$ labels the different supermultiplets,
and ${\rm PE}[i_k]$ is the plethystic exponential of the single-letter index of the $k$-th multiplet.
The pletyhstic exponential,
\be
{\rm PE}[i_k(t, y, U, V)] \equiv \exp\left\{ \sum_{m=1}^{\infty}\frac{1}{m}i_k (t^{m}, y^m, V^m)\chi_{{\mathcal R}_{k}}(U^{m}, V^m)\right\}\,,
\ee
implements the combinatorics of  symmetrization of the single letters, see {\it e.g.} \cite{Feng:2007ur, Benvenuti:2006qr, Aharony:2003sx}.
 As usual,  one can gauge fix the integral
over the gauge group and reduce it to an integral over the maximal torus, with the usual extra factor arising of van der Monde determinant.

In the following we focus on quiver gauge theories. The gauge group
will be taken to be a product of $SU(N)$ factors,
with the chiral matter
transforming in bifundamental representations. The gauge characters factorize into products
of fundamental and anti-fundamental characters of the relevant factors,
 $\chi_{{\mathcal R}_{a \bar b}}(U^{m}) \to \mbox{tr}(u_{a}^{m})\mbox{tr}\left(u_{b}^{\dagger m}\right)$.
For $SU(N)$  the adjoint character is
 $\chi_{adj}(U^{m})\equiv\mbox{tr}(u_{a}^{m})\mbox{tr}(u_{a}^{\dagger m})-1$.

The multi-letter contribution to the index of a chiral multiplet  (the plethystic exponential
of its single-letter index) can be elegantly written as a product of elliptic Gamma
functions~\cite{Dolan:2008qi}.
For a chiral superfield in
the bifundamental  representation $\Box \overline \Box$  of $SU(N_1) \times SU(N_2)$, and with IR R-charge equal to $r$,
one has
\bea \label{chiralelliptic}
 {\rm PE}[i_r(t, y, U)]& \equiv& \prod_{i=1}^{N_1} \prod_{j=1}^{N_2}\Gamma(t^{3r}\,z_i  w_j^{-1} ;\,t^3y,t^3/y),\qquad\\
\Gamma(z;p,q)& \equiv & \prod_{k,m=1}^\infty
\frac{1-p^{k+1}q^{m+1}/z}{1-p^{k}q^{m}\,z}\, .\nonumber
\eea
Here   $\{ z_k \}$,  $k=1, \dots N_1 \}$, and  $\{ w_k \}$, $k=1, \dots N_2 \}$, are complex numbers of unit modulus, obeying
$\prod_{k=1}^{N_1} z_k= \prod_{k=1}^{N_2} w_k=1$, which
parametrize the Cartan subalgebras of $SU(N_1)$ and $SU(N_2)$.

Similarly, the multi-letter contribution of a vector multiplet
in the adjoint of $SU(N)$ combines with the $SU(N)$ Haar measure to give
 the compact expression \cite{Dolan:2008qi, Gadde:2009kb}
\be \label{vectorelliptic}
\frac{\kappa^{N-1}}{N!}\oint_{\mathbb{T}_{n-1}}\prod_{i=1}^{N-1}
\frac{dz_{i}}{2\pi i\,z_{i}}\,\prod_{k\neq
\ell}\frac{1}{\Gamma(z_k/z_\ell;p,q)}\,\dots\,. \ee
The dots indicate that this is to be understood as a building block of the full matrix integral.
Here and everywhere the parameters $p$ and $q$ and $\kappa$ are taken to be
\be
p \equiv t^3 y  \, ,\quad q\equiv t^3/y\, , \quad \kappa \equiv (p;p)(q;q) 
\ee where $(a;b)\equiv\prod_{k=0}^\infty(1-ab^k)$. We will often
leave implicit the $q$ and $p$ dependence of the elliptic gamma
functions, $\Gamma(z; p,q) \to \Gamma(z)$.

\section{A universal result about ${\cal N}=2 \to {\cal N}=1$ flows}

Consider an ${\cal N}=2$ gauge theory where all the gauge couplings
are exactly marginal. Upon turning
on a mass term for the adjoint chiral multiplet inside the ${\cal N}=2$ vector multiplet, supersymmetry
is broken to ${\cal N}=1$ and the theory flows in the IR to an ${\cal N}=1$ superconformal
field theory with a quartic superpotential. The simplest example is  the flow between the ${\cal N}=2$ $\mathbb{Z}_2$ orbifold of ${\cal N}=4$
and the Klebanov-Witten theory. A large class of examples have been discussed in  \cite{Benini:2009mz}. 
For this general class of flows, there is a   universal linear
relation between the $a$ and $c$ conformal anomaly coefficients of the UV and IR theories \cite{Tachikawa:2009tt}.

It turns out that the superconformal indices of the UV and IR theories are also related in a simple
universal way, namely
\be \label{claim}
{\cal I}^{ {\cal N}=1}_{IR} (t, y) = {\cal I}^{{\cal N}=2}_{UV} (t, y, v=t) \,.
\ee
Choosing for definiteness the right-handed index, the definition of  the ${\cal N}=2$ superconformal index is
\be
{\cal I}^{{\cal N}=2} \equiv {\rm Tr} \, (-1)^F t^{2 (E + j_2)} y^{2 j_1} v^{-(r_{{\cal N} =2} + R)} \, ,
\ee
where $R$ and $r_{{\cal N}=2}$ are the quantum numbers under the $SU(2)_R \times U(1)_r$ R-symmetry.\footnote{
In our conventions, the bottom component $\phi$ of the ${\cal N}=2$ vector multiplet has $r_{{\cal N}=2} = -1$
(and of course $R=0$), while the scalar doublet in the hypermultiplet has $r_{{\cal N} =2} =0$ and $R = \pm 1/2$.}
The ${\cal N}=1$ and ${\cal N}=2$ R-symmetry quantum numbers are related as
\begin{equation}
r_{{\cal N}=1}=\frac{2}{3}(2R_{{\cal N}=2}-r_{{\cal N}=2})\,.
\end{equation}
Our claim is easily proved by recalling the single-letter indices of the ${\cal N}=2$ vector multiplet
and of the chiral multiplet (half-hypermultiplet), see {\it e.g.} \cite{Gadde:2009dj} 
\bea
i^{{\cal N} =2}_V(t, y, v)  & = &  \frac{t^2 v - \frac{t^4}{v} - t^3 (y + y^{-1}) + 2 t^6}{(1-t^3 y) (1 - t^3 y^{-1})}\\
i^{{\cal N} =2}_\chi(t, y, v)  & = & \frac{   \frac{t^2}{\sqrt{v}} - t^4 \sqrt{v} } {(1-t^3 y) (1 - t^3 y^{-1})}\,.
\eea
Comparing with (\ref{iC1}) and  (\ref{iV1}), we see that
\bea
i^{{\cal N} =2}_V(t,y, v=t)  & = & i^{{\cal N} =1}_V(t,y)\\
i^{{\cal N} =2}_\chi(t,y, v=t)  & = & i^{{\cal N} =1}_{\chi(r=\frac{1}{2}) }(t,y)\,.
\eea
So setting $v=t$ has the effect of converting the ${\cal N}=2$ vector multiplet to ${\cal N}=1$ vector
multiplets, and 
of changing the R-charge of the chiral multiplets from $r_{{\cal N}=1}=2/3$ to  $r_{{\cal N}=1}=1/2$,
which is the correct IR value since a quartic superpotential is generated from the decoupling of the
adjoint chiral multiplets. Since both the conformal anomaly coefficients and the index undergo
a universal transformation between the UV and IR of this class of RG flows, one may wonder whether
there is any simple connection between the index and the anomaly coefficients.

\section{The $Y^{p,q}$ quiver gauge theories}\label{Ypqsec}

Let us begin by recalling the basic facts about the $Y^{p,q}$ quiver gauge theories \cite{Benvenuti:2004dy}.
 The  fields are of four types: $U_{\a=1,2}$, $V_{\a=1,2}$, $Y$ and $Z$.
There are $2p$ gauge groups, and $4p+2q$ bifundamental fields: $p$ fields of type $U$, $q$ fields of type $V$, $p-q$ fields of type $Z$, and $p+q$ fields
of type $Y$. The $Y^{p,q}$ quiver diagram is obtained by a recursive procedure starting with
$Y^{p,p}$,
which is a familiar $\mathbb{Z}_{2p}$ orbifold
of ${\cal N}=4$ SYM. The superpotential takes the form
\be\label{YPqsup}
\mathbf{W}=\sum \e^{\a\b}\Tr\left(U^k_\a\, V^k_\b\, Y^{2k+2}+V^k_\a\, U^{k+1}_\b\, Y^{2k+3}\right)+\e_{\a\b}\sum \Tr\left(
 Z^k\, U^{k+1}_\a\, Y^{2k+3}\, U^k_\b\right)\, ,\nonumber
\ee
where the cubic and quartic gauge-invariant terms are read off from the quiver diagram.
There are $2q$ terms in the first sum and $p-q$ terms in the second sum. For the Klebanov-Witten theory, $T^{1,1}=Y^{1,0}$ 
has only quartic terms.

\begin{figure}
\begin{centering}
\includegraphics[scale=0.4]{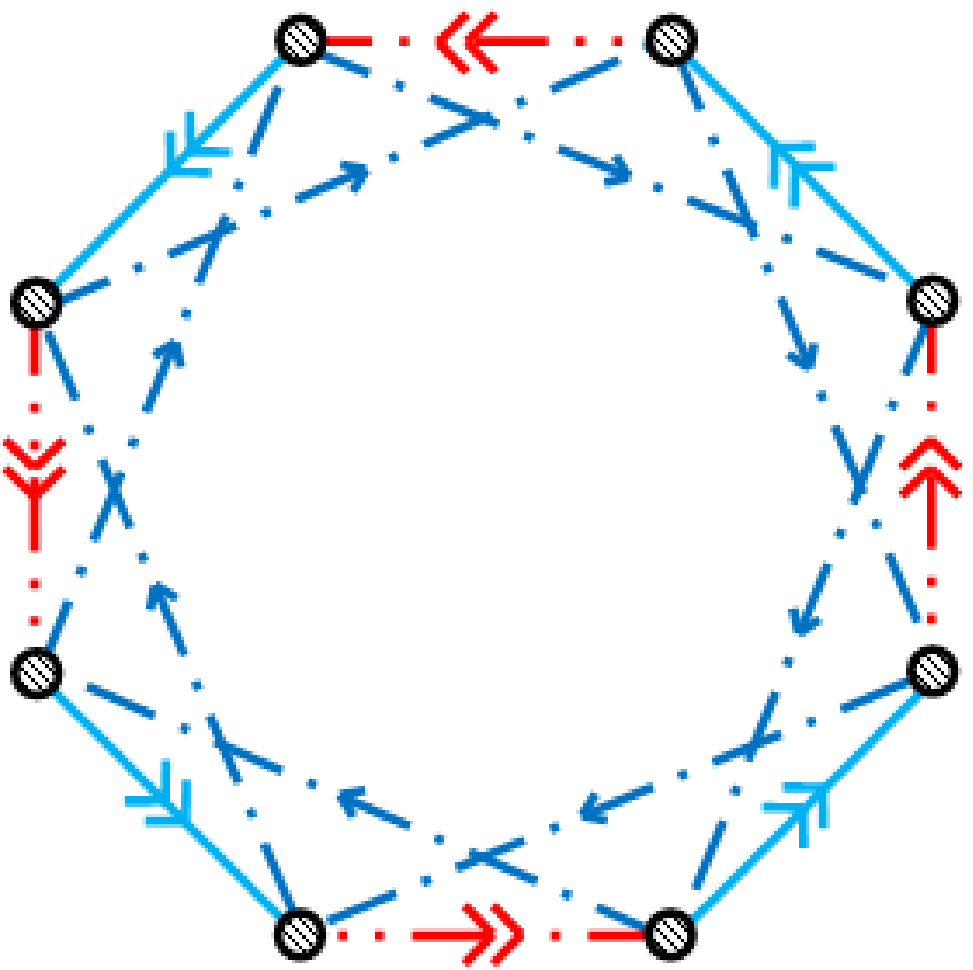} \qquad \qquad  \includegraphics[scale=0.4]{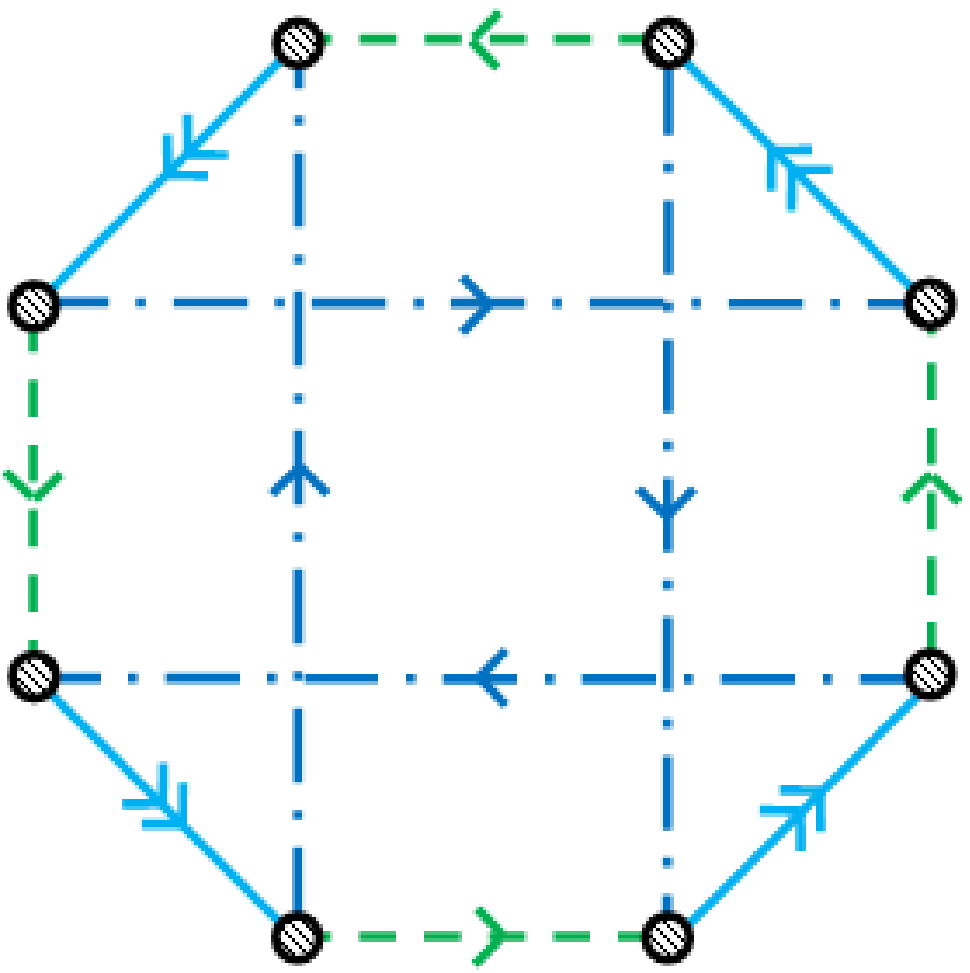}
\par\end{centering}
\caption{\label{Y44Y40} Left: quiver diagram for $Y^{4,4}$. Right: quiver diagram for  $Y^{4,0}$.}
\end{figure}

The R-charges are determined as follows \cite{Benvenuti:2004dy, Yamazaki:2008bt}.
Requiring the vanishing of the NSVZ beta functions and that each term of the superpotential
has R-charge 2,  the R-charges of all the fields are fixed in terms of two independent parameters $x$ and $y$,
\be
r_{Z^k}=x,\qquad r_{Y^k}=y,\qquad r_{U^k_\a}=1-\half(x+y),\qquad r_{V^k_\a}=1+\half(x-y)\,.
\ee
This twofold ambiguity is related to the existence of two $U(1)$ global symmetries, and
is resolved  by  $a$-maximization. One finds~\cite{Benvenuti:2004dy}
\be
y_{p,q}&=&\frac{1}{3q^2}\left\{-4p^2+2pq+3q^2+(2p-q)\sqrt{4p^2-3q^2}\right\}\,,\\
x_{p,q}&=&\frac{1}{3q^2}\left\{-4p^2-2pq+3q^2+(2p+q)\sqrt{4p^2-3q^2}\right\}\,.\nonumber
\ee
For any $p$, there are simple special cases. The
 $Y^{p,p}$ quiver corresponds to the $\mathbb{Z}_{2p}$ orbifold of $\mathbb{C}^3$. In this case all the superpotential
terms are cubic, the theory is  exactly conformal and all R-charges are equal to $\frac{2}{3}$.
This theory has ${\mathcal N}=1$ supersymmetry
for general $p$ while for $p=1$ the supersymmetry is enhanced to $\mathcal{N}=2$.
At the other extreme,  the $Y^{p,0}$ quiver corresponds to   a $\mathbb{Z}_p$ orbifold of the conifold.  All the R-charges are equal to $\half$ and the superpotential
is quartic.
The associated quiver diagrams for $p=4$ are shown in figure \ref{Y44Y40}.

The  charges of the fields under the global  symmetries $U(1)_B$, $U(1)_s$ and $SU(2)_l$  
and the color-coding of the arrows are indicated below.
\be \label{charges}
\begin{tabular}{|c|c|c|c|c|}
\hline
&$U(1)_B$&$U(1)_s$&$SU(2)_l$&Arrows\\
\hline
$U$&$-p$ &$0$&$\pm\half$&\includegraphics[scale=0.5]{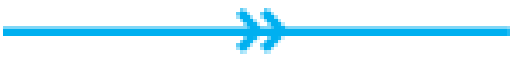}\\
$V$&$q$&$\frac{1}{2}$&$\pm\half$&\includegraphics[scale=0.5]{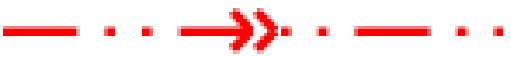}\\
$Z$&$p+q$&$\frac{1}{2}$&$0$&\includegraphics[scale=0.5]{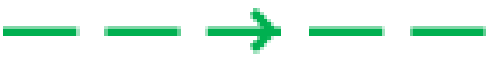}\\
$Y$&$p-q$&$-\frac{1}{2}$&$0$&\includegraphics[scale=0.5]{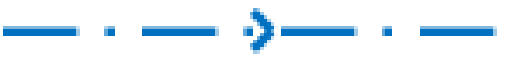}\\
\hline
\end{tabular}
\ee
We can refine the  index by chemical potentials for the global symmetries,
\bea
{\mathcal I}^{\tt L}(t,y,a,b,h)  & = & \Tr (-1)^F\, t^{3 (2 j_1 -r) }\,y^{2j_2}\,a^{2s}\,b^{2l} \,h^{Q_B} \, \\
{\mathcal I}^{\tt R}(t,y,a,b,h)  & = & \Tr (-1)^F\, t^{ 3 (2j_2 + r) }\,y^{2j_1}\,a^{2s}\,b^{2l} \,h^{Q_B}\,.
\eea
In practice we can focus on say the left-handed index.  The right-handed index of a given
theory is obtained from the left-handed index of the same theory
by conjugation of the flavor quantum numbers, $a \to 1/a$, $h \to 1/h$.

Given a $Y^{p,q}$ quiver diagram, it is immediate to combine
 the chiral and vector building blocks (\ref{chiralelliptic}), (\ref{vectorelliptic})
 and construct the matrix integral that calculates the corresponding index.  We illustrate the procedure
in the two simplest examples.

\subsection*
{$\bullet$ $Y^{1,0}$ ($T^{1,1}$)}

\begin{figure}
\begin{centering}
\includegraphics[scale=0.6]{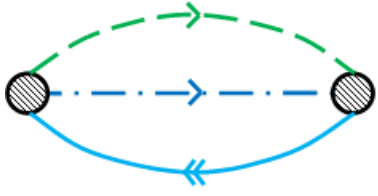}
\par\end{centering}

\caption{\label{fig:Y10} Quiver diagram for $Y^{1,0}$ (the conifold theory  $T^{1,1}$). The solid
(cyan) arrow represents the $U$ field, the dash-dot (blue) arrow  represents the
$Y$ field and the dashed (green) arrow represents the $Z$ field.}
\end{figure}

The quiver of $T^{1,1}$ is shown in figure \ref{fig:Y10}. The index
can be simply read from the quiver diagram,
\begin{equation}
  \begin{split}
  \mathcal{I}_{1,0}
  =&\prod_{k=1}^2\left[\frac{\kappa^{N-1}}{N!}
   \oint_{\mathbb{T}}\prod^{N-1}_{i=1}\frac{dz^{(k)}_i}{2\pi iz^{(k)}_i}\frac{1}{\prod_{i\neq
   j}\Gamma(z^{(k)}_i/z^{(k)}_j)}\right]\\
   &\times\prod^N_{i,j=1}\Gamma(t^{3 r_U}b^\pm z^{(2)}_i/z^{(1)}_j)
   \prod^N_{i,j=1}\Gamma(t^{3r _Y}a^{-1}z^{(1)}_i/z^{(2)}_j)\prod^N_{i,j=1}\Gamma(t^{3 r_Z}a \,z^{(1)}_i/z^{(2)}_j)
   \end{split}
\end{equation}
where the
R-charges are
\begin{equation}
  r_{U}=r_Y=r_Z=\frac{1}{2}.
\end{equation}
The fact that $Y$ and $Z$ share the same R-charge leads to the symmetry enhancement $U(1)_s \to SU(2)_s$.

\subsection*
{$\bullet$ $Y^{1,1}$ ($\mathbb{C}^{2}/\mathbb{Z}_{2}\times\mathbb{C}$)}

\begin{figure}
\begin{centering}
\includegraphics[scale=0.6]{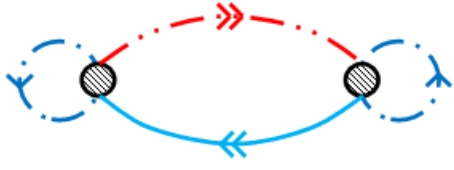}
\par\end{centering}

\caption{\label{fig:Y11}Quiver of $Y^{1,1}$ theory. Solid (cyan)
arrow represents $U$ field, dash-dot-dot arrow (red) represents $V$
field and dash-dot arrow (blue) represents $Y$ field.}
\end{figure}

The quiver corresponding to $Y^{1,1}$ is shown in figure
\ref{fig:Y11}. This theory is the familiar $\mathbb{Z}_{2}$ orbifold of ${\cal
N}=4$ SYM which in fact preserves ${\cal N}=2$ supersymmetry, but
we write its ${\cal N}=1$ index for a uniform analysis,\footnote{The index for this theory has been already calculated at large $N$ \cite{Nakayama:2005mf, Gadde:2009dj}.}
\begin{equation}
  \begin{split}
    \mathcal{I}_{1,1}
    =&\prod_{k=1}^2\left[\frac{ \kappa^{N-1}}{N!}
     \oint_{\mathbb{T}}\prod^{N-1}_{i=1}\frac{dz^{(k)}_i}{2\pi iz^{(k)}_i}\frac{1}{\prod_{i\neq
     j}\Gamma(z^{(k)}_i/z^{(k)}_j)}\right]\\&\times
     \left[\prod^N_{i,j=1}\Gamma(t^{3 r _U}b^\pm z^{(2)}_i/z^{(1)}_j)
     \prod^N_{i,j=1}\Gamma(t^{3 r_V}b^\pm a\, z^{(1)}_i/z^{(2)}_j)\right]\\&
     \times\left[\prod^N_{i\neq j}\Gamma(t^{3 r_Y} a^{-1} z^{(1)}_i/z^{(1)}_j)\Gamma(t^{3 r_Y})^{N-1}
     \prod^N_{i\neq j}\Gamma(t^{3 r_Y} a^{-1} z^{(2)}_i/z^{(2)}_j)\Gamma(t^{3 r_Y})^{N-1}\right]\,.
  \end{split}
\end{equation}

\subsection {Toric Duality}

\begin{figure}
\begin{centering}
\includegraphics[scale=0.5]{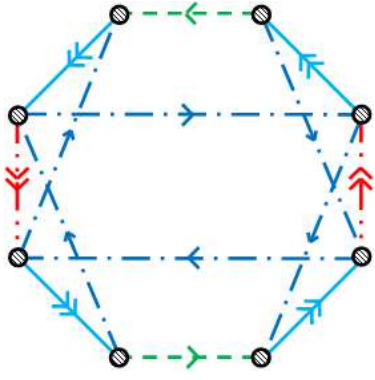}
\par\end{centering}

\caption{\label{fig:Y42_1}Quiver diagram for $Y^{4,2}$, obtained from
$Y^{4,4}$ by using the procedure in \cite{Benvenuti:2004dy}.}
\end{figure}

\begin{figure}
\begin{centering}
\includegraphics[scale=0.5]{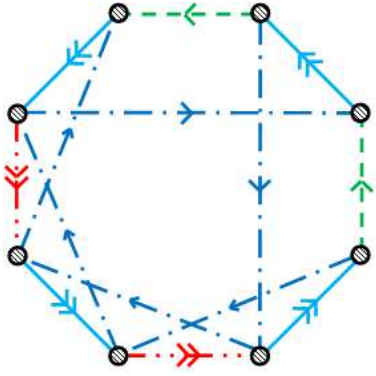}
\par\end{centering}

\caption{\label{fig:Y42_2}A different quiver diagram for $Y^{4,2}$, related to the diagram above by toric duality.}
\end{figure}

A toric Calabi-Yau singularity
may have several equivalent quiver representations, related by what has
been called ``toric duality'' \cite{Feng:2000mi}. In terms of the gauge theories on D3 branes
probing the singularity, two toric-dual quiver diagrams define two UV theories that flow
to the  same IR superconformal fixed point. Toric duality can in fact be understood in terms of the usual Seiberg duality
of super QCD \cite{Beasley:2001zp,Feng:2001bn,Ooguri:1997ih,Cachazo:2001sg,Feng:2001xr,Feng:2002zw,
Franco:2003ja}.
In particular, the prescription  \cite{Benvenuti:2004dy}
for finding the quiver theory associated $Y^{p,q}$ 
does not lead to unique answer, rather to a family of quivers
related by toric duality. The simplest example occurs for $Y^{4,2}$: the  pair of toric-dual quivers
 associated to $Y^{4,2}$ is shown  in figures
\ref{fig:Y42_1} and  \ref{fig:Y42_2}.

We are now going to check the equality of the indices of  two dual theories using
an identity between elliptic hypergeometric integrals.

\begin{figure}
\begin{centering}
\includegraphics[scale=0.5]{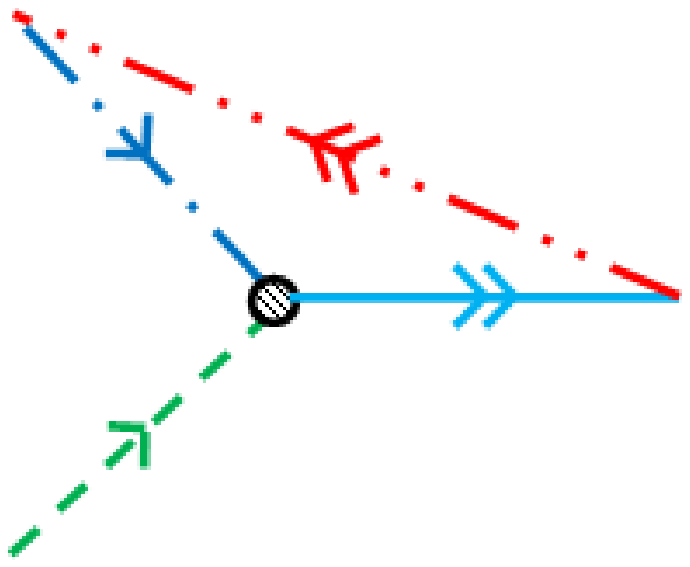}$\phantom{O}$\includegraphics[scale=0.5]{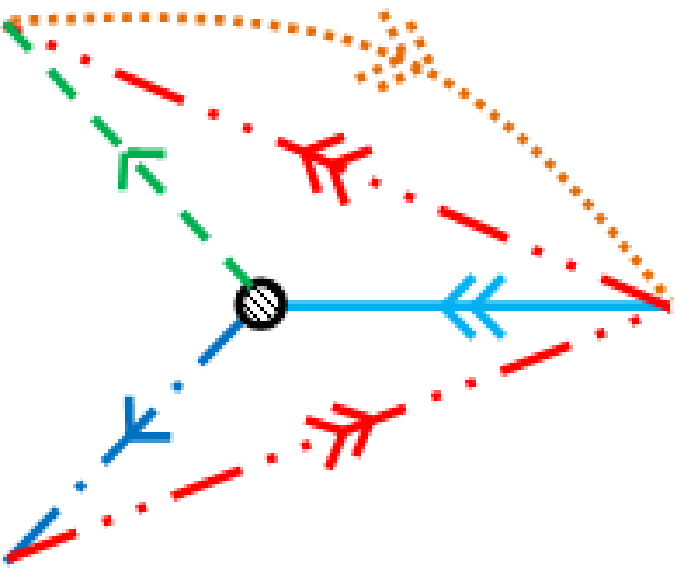}$\phantom{O}$\includegraphics[scale=0.5]{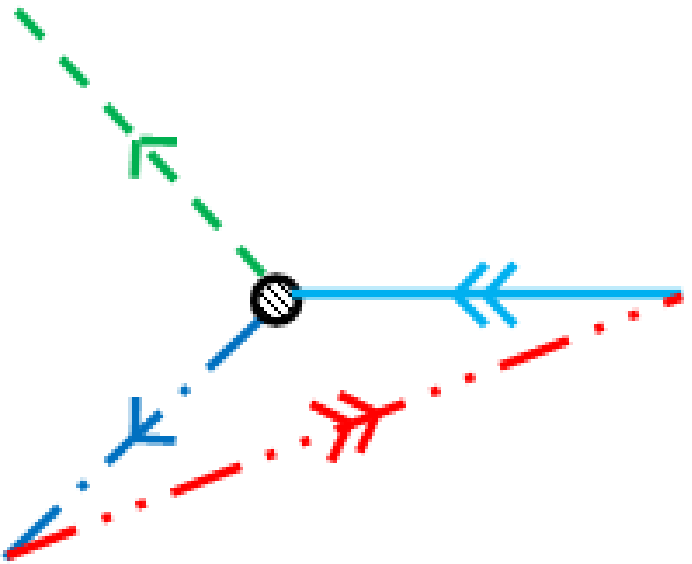}
\par\end{centering}
\caption{\label{fig:SeibergDual}Left: initial quiver. The node represents a $SU(N)$ gauge
group. The effective number of flavors is $N_f = 2 N$.
Middle: quiver after Seiberg duality. The node represents the Seiberg dual gauge group $SU(2N-N) = SU(N)$.
All arrows are reversed and  mesons (with appropriate R-charges) are added. Right: the dash-dot-dot (red) and dot (orange)
mesons cancel each other out by equ.(4.15). This can be understood physically in terms of integrating out massive degrees of freedom [31].}
\end{figure}

Consider the $k$-th node of a $Y^{p,q}$ quiver with one incoming $Y$, one incoming $Z$ and
an outgoing $U$ doublet (see the first diagram in figure \ref{fig:SeibergDual}).
Its contribution to the index (suppressing global symmetry charges) is
\begin{equation}
  \begin{split}
    \mathcal{I}^k_{p,q}=&
    \frac{\kappa^{N-1}}{N!}
     \oint_{\mathbb{T}}\prod^{N-1}_{i=1}\frac{dz^{(k)}_i}{2\pi iz^{(k)}_i}\frac{1}{\prod_{i\neq
     j}\Gamma(z^{(k)}_i/z^{(k)}_j)}\\
     &\times\prod_{i,j}\Gamma(t^{3r_Z}\frac{z^k_i}{z^Z_j})
     \prod_{i,j}\Gamma(t^{3r_Y}\frac{z^k_i}{z^Y_j})
     \prod_{i,j}\Gamma(t^{3r_U}\frac{z^U_j}{z^k_i})^2
     \prod_{i,j}\Gamma(t^{3r_V}\frac{z^Z_i}{z^U_j})^2.
  \end{split}
\end{equation}
where $z^U$, $z^Y$ and $z^Z$ represents the "flavor" group of $U$, $Y$ and $Z$.
This is precisely the $A_n$-type integral defined in \cite{rains},
\begin{equation}
  \begin{split}
    \mathcal{I}^k_{p,q}=I^{(N-1)}_{A_{N-1}}(Z|t^{3r_Z}/z^Z_j,t^{3r_Y}/z^Y_j;t^{3r_U}z^U_j,t^{3r_U}z^U_j;p,q)
    \prod_{i,j}\Gamma(t^{3r_V}\frac{z^Z_i}{z^U_j})^2 \,.
  \end{split}
\end{equation}
This integral obeys the balancing condition
\begin{equation}
  \prod_{j=1}^N\frac{t^{3r_Z}}{z^Z_j}\frac{t^{3r_Y}}{z^Y_j}t^{3r_U}z^U_jt^{3r_U}z^U_j=(pq)^{N}\, ,
\end{equation}
thanks to the relation
\begin{equation}
  r_Y+r_Z+2 r_U=y_{p,q}+x_{p,q}+2[1-\frac{1}{2}(x_{p,q}+y_{p,q})]=2 \,.
\end{equation}
Then the following identity holds \cite{rains}:
\begin{equation}
  I^{(m)}_{A_n}(Z|t_i\ldots,u_i\ldots)=\prod_{r,s=1}^{m+n+2}\Gamma(t_ru_s)I^{(n)}_{A_m}
  (Z|\frac{T^{\frac{1}{m+1}}}{t_i}\ldots,\frac{U^{\frac{1}{m+1}}}{u_i}\ldots).
\end{equation}
So we have
\begin{equation}
  \begin{split}
    \mathcal{I}^k_{p,q}=&I^{(N-1)}_{A_{N-1}}(Z|t^{3r_Z}/z^Z_j,t^{3r_Y}/z^Y_j;t^{3r_U}z^U_j,t^{3r_U}z^U_j;p,q)\prod_{i,j}\Gamma(t^{3r_V}\frac{z^Z_i}{z^U_j})^2\\
    =&\prod_{i,j=1}^N\Gamma(t^{3(r_Z+r_U)}\frac{z^U_i}{z^Z_j})^2\prod_{i,j=1}^N\Gamma(t^{3(r_Y+r_U)}\frac{z^U_i}{z^Y_j})^2\\
     &\times I^{(N-1)}_{A_{N-1}}(Z|t^{3r_Y}z^Z_j,t^{3r_Z}z^Y_j;t^{3r_U}/z^U_j,t^{3r_U}/z^U_j;p,q)
      \prod_{i,j=1}^N\Gamma(t^{3r_V}\frac{z^Z_j}{z^U_i})^2\\
    =&\prod_{i,j=1}^N\Gamma(t^{3r_V}\frac{z^U_i}{z^Y_j})^2
      I^{(N-1)}_{A_{N-1}}(Z|t^{3r_Y}z^Z_j,t^{3r_Z}z^Y_j;t^{3r_U}/z^U_j,t^{3r_U}/z^U_j;p,q),
  \end{split}
\end{equation}
where we have  used $r_V = r_Z+r_U$ and
\be
\label{eq:cancellation}
\Gamma(t^{3r_V}\frac{z^Z_i}{z^U_j})\Gamma(t^{3(r_Y+r_U)}\frac{z^U_j}{z^Z_i}) = 1 \,.
\ee

\begin{figure}
\begin{centering}
\includegraphics[scale=0.4]{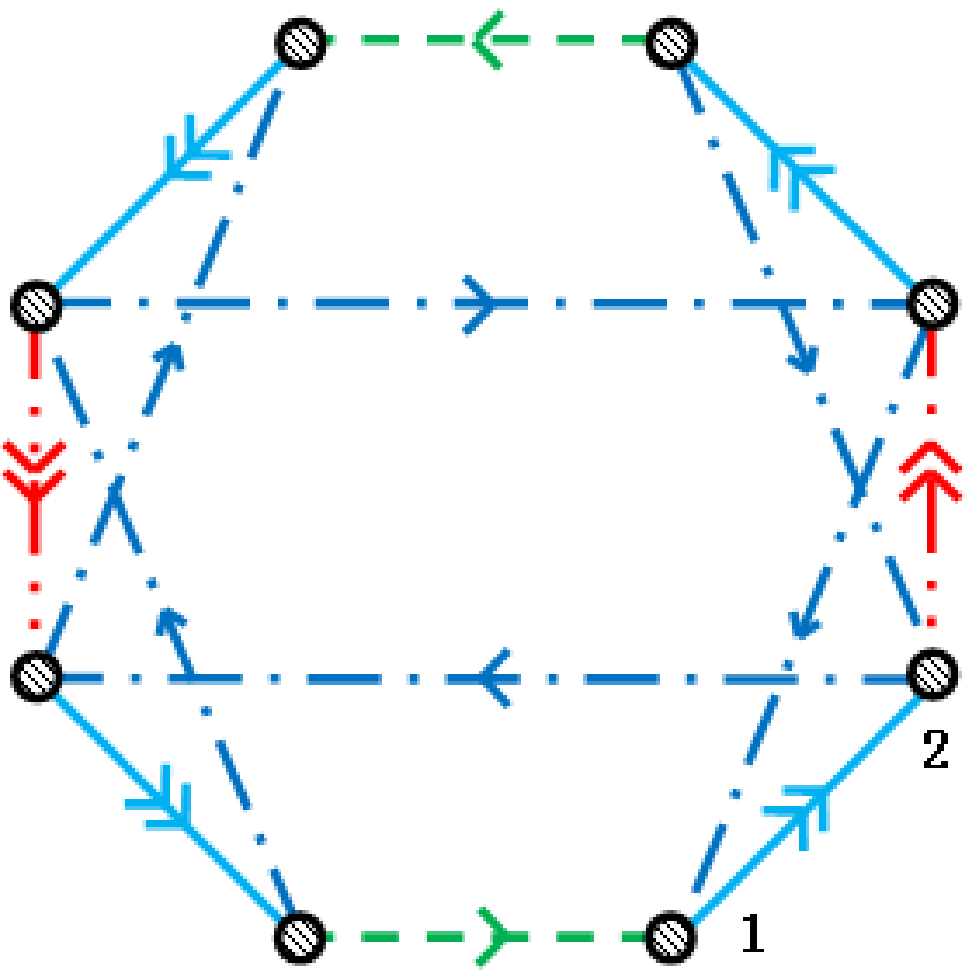}\phantom{OO}\includegraphics[scale=0.4]{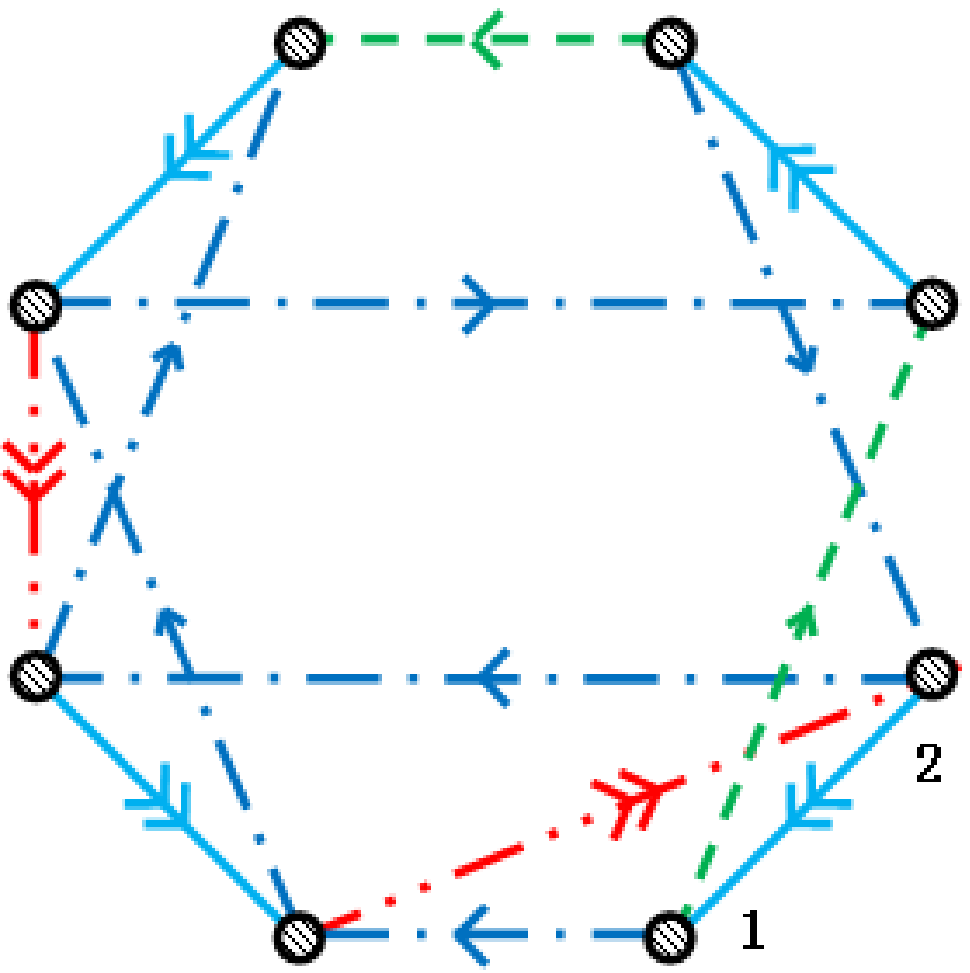}\phantom{OO}\includegraphics[scale=0.4]{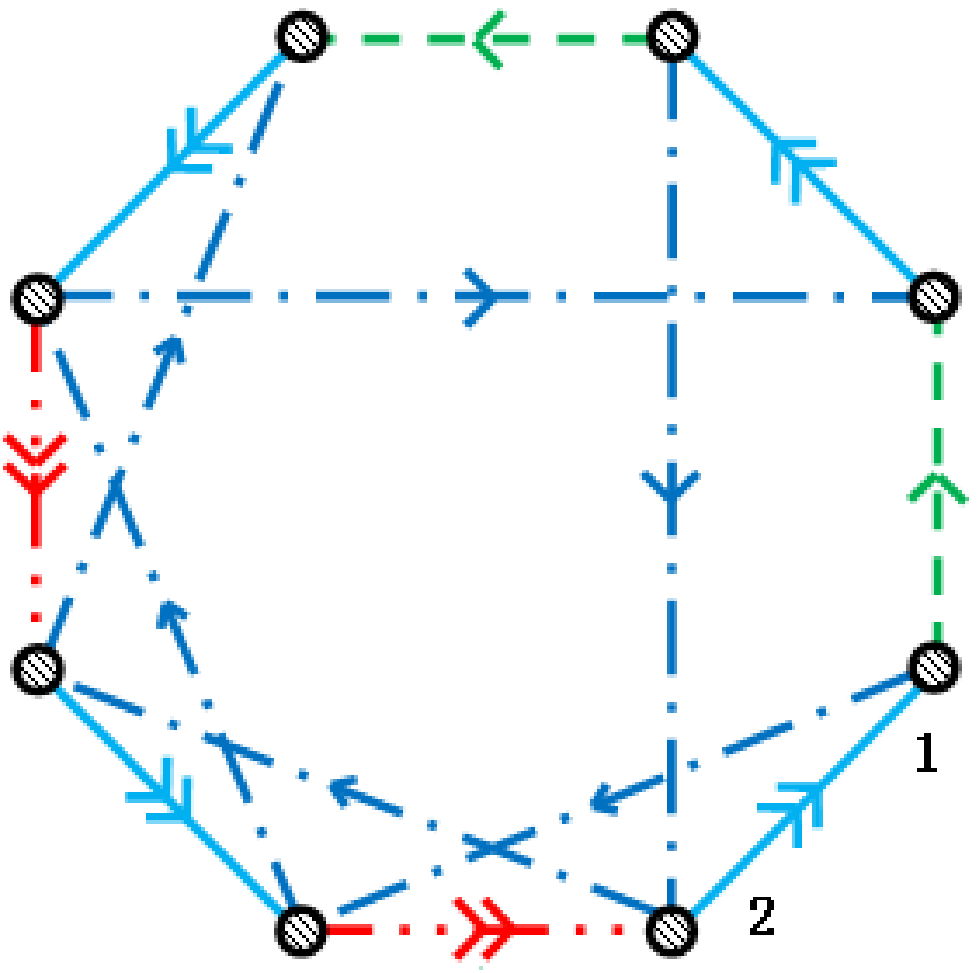}
\par\end{centering}

\caption{\label{fig:DualityInY42}Left: $Y^{4,2}$ quiver in Figure 3.
Middle: Seiberg dual on node $1$. Right: the quiver in Figure 4
 is obtained by swap node $1$ and $2$ in the middle
figure.}
\end{figure}

For example, one can perform this duality on one of the $YZ\bar{U}$ nodes of the $Y^{4,2}$ quiver in
figure \ref{fig:Y42_1} and obtain the quiver in figure
\ref{fig:Y42_2}. The procedure is illustrated in figure
\ref{fig:DualityInY42}.

This transformation can be represented on a quiver as a local graph transformation of figure \ref{fig:SeibergDual}. It has
the interpretation of Seiberg duality on the node. (In fact the same elliptic hypergeometric identity was used in \cite{Dolan:2008qi}
to demonstrate the equality of the index under Seiberg duality.)
Iterating this step, we can reach all the toric phases of any $Y^{p,q}$ gauge theory.

\section{Large $N$ evaluation of the index}
\label{largeNsec}

In the large $N$ limit the leading contribution to the index is evaluated  using matrix models techniques (see e.g.~\cite{Aharony:2003sx,Kinney:2005ej}).
 Let
$\{e^{\alpha_{ai}}\}_{i=1}^{N_{a}}$ denote the $N_{a}$ eigenvalues
of $u_{a}$. Then the matrix model integral (\ref{integral}) is,
\begin{equation}
{\cal I}(x)=\int\prod_{a,i}[d\alpha_{ai}]\exp\left\{ -\sum_{ai\neq bj}V_{b}^{a}(\alpha_{ai}-\alpha_{bj})\right\}\,.
\end{equation}
Here, the potential $V$ is the following function
\begin{equation}
V_{b}^{a}(\theta)=\delta_{b}^{a}(\ln2)+\sum_{n=1}^{\infty}\frac{1}{n}[\delta_{b}^{a}-i_{b}^{a}(x^{n})]\cos n\theta\,,
\end{equation}
where, $i_{b}^{a}(x)$ is the total single letter index in the representation
$r^{a}\otimes r_{b}$. Writing the density of the eigenvalues $\{e^{\alpha_{ai}}\}$
at the point $\theta$ on the circle as $\rho_{a}(\theta)$, we reduce
it to the functional integral problem,
\begin{equation}
{\cal I}(x)=\int\prod_{a}[d\rho_{a}]\exp\{-\int d\theta_{1}d\theta_{2}\sum_{a,b}n_{a}n_{b}\rho_{a}(\theta_{1})V_{b}^{a}(\theta_{1}-\theta_{2})\rho_{b}^{\dagger}(\theta_{2})\}\,.
\end{equation}
For large $N$, we can evaluate this expression with the saddle point
approximation,\begin{eqnarray*}
{\cal I}(x) & = &  \prod_{k}\frac{1}{\det(1-i(x^{k}))}\,.\end{eqnarray*}
For $SU(N)$ gauge groups instead of $U(N)$, the result is modified
as follows,
\begin{equation}
{\cal I}(x)=\prod_{k}\frac{e^{-\frac{1}{k}\mbox{tr }i(x^{k})}}{\det(1-i(x^{k}))}\,.
\end{equation}
 Here  $i(x)$ is the  matrix with entries $i_{b}^{a}(x)$.
We will see examples of such matrices below.

The single-trace partition function can be obtained from the
full partition function,
\begin{eqnarray}
{\cal I}_{s.t.} & = & \sum_{n=1}^{\infty}\frac{\mu(n)}{n}\log{\cal I}(x^{n})\\
 & = & -\sum_{k=1}^{\infty}\frac{\varphi(k)}{k}\log[\det(1-i(x^{k}))]-\sum_{n=1}^{\infty}\frac{\mu(n)}{n}\sum_{k=1}^{\infty} \frac{\mbox{tr }i(x^{n k})}{k}\\
 & = & -\sum_{k=1}^{\infty}\frac{\varphi(k)}{k}\log[\det(1-i(x^{k}))]-\mbox{tr }i(x)\,.
 \end{eqnarray}
The second term in the summation would be absent for the $U(N)$ gauge theories.
Here $\mu(n)$ is the M\"obius function ($\mu(1)\equiv1,\:\mu(n)\equiv0$
if $n$ has repeated prime factors and $\mu(n)\equiv(-1)^{k}$ if
$n$ is the product of $k$ distinct primes) and $\varphi(n)$ is the
Euler Phi function, defined as the number of positive integers less
than $n$ that are coprime to $n$. We have used the properties
\begin{equation}
\sum_{d|n}d\mu(\frac{n}{d})=\varphi (n),\qquad\qquad\sum_{d | n}\mu(d)=\delta_{n,1}.
\end{equation}

After deriving the general expression for the superconformal index
of a quiver gauge theory let us study some concrete examples. Recall the single-letter indices
\begin{eqnarray}
i_{V}(t,y) & = & \frac{2t^{6}-t^{3}(y+\frac{1}{y})}{(1-t^{3}y)(1-t^{3}y^{-1})}\,,\\
i_{\bar \chi(r)}(t,y) & = & \frac{t^{3r}}{(1-t^{3}y)(1-t^{3}y^{-1})}\,\quad
i_{\chi(r)}(t,y)  =  -\frac{t^{3(2-r)}}{(1-t^{3}y)(1-t^{3}y^{-1})}\, ,
\end{eqnarray}
where for future convenience we have split the index of the matter multiplet into a chiral and an antichiral contribution.
Let us write down explicit expressions for the index in some examples

\subsection*
{$\bullet$ $Y^{1,0}\, (T^{1,1})$}
For the conifold gauge theory, $U(1)_s$ is enhanced to $SU(2)_s$ so the global symmetry is  $SU(2)_s\times SU(2)_l$. Assigning the chemical potentials $a$ and $b$, for the two $SU(2)$s,
the single letter
index matrix $i_{1,0}(t,y)$ is
\begin{equation}
i_{1,0}=\left(\begin{array}{cc}
i_{V} & (a+\frac{1}{a})(i_{\chi(\frac{1}{2})}+i_{\bar\chi(\frac{1}{2})})\\
(b+\frac{1}{b})(i_{\chi(\frac{1}{2})}+i_{\bar\chi(\frac{1}{2})}) &
i_{V}\end{array}\right)\,,
\end{equation}
and the single-trace index evaluates to
\begin{eqnarray}\label{indT11}
\label{eq:singleTracePartitionFunction:largeN}
{\cal I}_{s.t.} & = & -\sum_{k=1}^{\infty}\frac{\varphi(k)}{k}\log[(1-i_{V}(x^{k}))^{2}-(a+\frac{1}{a})(b+\frac{1}{b})(i_{\chi(\frac{1}{2})}(x^k)+i_{\bar \chi (\frac{1}{2})}(x^k))^{2}]-2 i_{V}(x)\nonumber\\
 & = & \frac{t^{3}ab}{1-t^{3}ab}+\frac{t^{3}
\frac{a}{b}}{1-t^{3}\frac{a}{b}}+\frac{t^{3}\frac{b}{a}}{1-t^{3}
\frac{b}{a}}+\frac{t^{3}\frac{1}{ab}}{1-t^{3}\frac{1}{ab}}\,.
 \end{eqnarray}
This is the index for the theory where {\it both} the overall and the relative $U(1)$ degrees of freedom have
been removed. The overall $U(1)$ is completely decoupled,  while the relative $U(1)$
has positive beta function and decouples in the IR. The removal of the relative
$U(1)$ introduces certain double-trace terms in the superpotential which are
important to achieve exact conformality~\cite{Dymarsky:2005nc}.
We have used the following property of Euler Phi function.
\begin{equation}
\sum_{k=1}^{\infty}\frac{\varphi(k)}{k}\log(1-x^{k})=\frac{-x}{1-x}.
\end{equation} We will match the expression~\eqref{indT11} to the gravity computation.

\subsection*
{$\bullet$ $Y^{1,1}\,(\mathbb{C}^{2}/\mathbb{Z}_{2}\times\mathbb{C})$}
The index for this theory was already obtained in \cite{Nakayama:2005mf, Gadde:2009dj}.
The single letter index matrix $i_{1,1}(t,y)$ is given by
\begin{equation}
i_{1,1}=\left(\begin{array}{cc}
i_{V}+a^{-1}i_{\chi(\frac{2}{3})}+a i_{\bar\chi(\frac{2}{3})} & (b+\frac{1}{b})(a i_{\chi(\frac{2}{3})}+i_{\bar\chi(\frac{2}{3})})\\
(b+\frac{1}{b})(i_{\chi(\frac{2}{3})}+a^{-1} i_{\bar\chi(\frac{2}{3})}) &
i_{V}+a^{-1}i_{\chi(\frac{2}{3})}+a\, i_{\bar\chi(\frac{2}{3})}\end{array}\right)\,,
\end{equation} and the index evaluates to
\begin{eqnarray}
{\cal I}_{s.t.} & = & -\sum_{k=1}^{\infty}\frac{\varphi(k)}{k}
\log[(1-i_{V}(x^{k})-a^{-1}i_{\chi(\frac{2}{3})}(x^{k})-a\, i_{\bar\chi(\frac{2}{3})}(x^{k}))^{2}-(b+\frac{1}{b})^2 \frac{1}{a}(a\, i_{\chi(\frac{2}{3})}(x^{k})+i_{\bar\chi(\frac{2}{3})}(x^{k}))^{2}] \nonumber\\
&& -2(i_{V}(x)+a^{-1} \, i_{\chi(\frac{2}{3})}(x)+ a\, i_{\bar\chi(\frac{2}{3})}(x))\nonumber\\
 & = & 2\frac{t^{2}a}{1-t^{2}a}+\frac{t^{4} b^2 a^{-1}}{1-t^{4}b^2 a^{-1}}+\frac{t^{4}b^{-2} a^{-1}}{1-t^{4}b^{-2} a^{-1}}-2 \frac{a\, t^{2}-a^{-1}\, t^{4}}{(1-t^{3}y)(1-t^{3}y^{-1})}\,.
 \end{eqnarray}
Again, we have subtracted both the overall and relative $U(1)$ degrees of freedom (in this case it is appropriate to subtract ${\cal N}=2$ vector
multiplets).

\subsection*
 {$\bullet$  General $Y^{p,q}$}

A simple generalization gives the index for $Y^{p,0}$ ($T^{1,1}/\mathbb{Z}_{p}$) and for $Y^{p,p}$ ($\mathbb{C}^3/\mathbb{Z}_{2p}$),
\be
Y^{p,p}\;:\qquad&&\det \left(1-i(t)\right)=\frac{(1-t^{4p})^2(1-t^{2p})^2}{(1-t^3 y)^{2p}(1-t^3 y^{-1})^{2p}}\,,\\
Y^{p,0}\;:\qquad&&\det \left(1-i(t)\right)=\frac{(1-t^{3p})^4}{(1-t^3 y)^{2p}(1-t^3 y^{-1})^{2p}}\,.\nonumber
\ee 
In fact the determinant of the adjacency
matrix appears to factorize for general $Y^{p,q}$,  to give\footnote{We have checked this result in several cases
but have not attempted an analytic proof.}
\be \det
\left(1-i(t)\right)=
\frac{\left[1-t^{3p\left(1+\half(x_{p,q}-y_{p,q})\right)}\right]^2\,\left[1-t^{3p+\frac{3q}{2}
\left(1-\half(x_{p,q}+y_{p,q})\right)}\right]^2}{(1-t^3 y)^{2p}(1-t^3 y^{-1})^{2p}}\,.
\ee
Thus the single-trace partition function is\footnote{
Curiously, this is exactly twice the index of the chiral mesons denoted ${\mathcal L}_+$ (first term) and
${\mathcal L}_-$ (second term) in~\cite{Benvenuti:2005cz}. We don't have a deeper understanding of this observation.
On the gravity sides, the chiral mesons of 
${\mathcal L}_{+/-}$ were identified in~\cite{Kihara:2005nt} (see also~\cite{Oota:2005mr} with
 the zero modes of the scalar Laplacian on the $Y^{p,q}$ manifold.). }
\be
{\mathcal I}_{p,q}^{s.p.}&=&2\left[\frac{t^{\frac{p (  3 q+2 p - \sqrt{4 p^2 - 3 q^2})}{q}}}{1-t^{\frac{p (2 p + 3 q - \sqrt{4 p^2 - 3 q^2})}{q}}}
+\frac{t^{\frac{p ( 3 q-2 p + \sqrt{4 p^2 - 3 q^2})}{q}}}{1-t^{\frac{p (2 p + 3 q - \sqrt{4 p^2 - 3 q^2})}{q}}}\right]\
\nonumber
\ee
Again, this is the result with all $U(1)$ factors subtracted.
  If one introduces a chemical potential $b^{2l}$ for the global $SU(2)_l$
and a chemical potential $a^{2s}$ for the global $U(1)_s$ of table~\ref{charges}  the  index becomes
\be
{\mathcal I}_{p,q}^{s.p.}&=&\frac{a^{-p} b^{p+q}t^{\frac{p (  3 q+2 p - \sqrt{4 p^2 - 3 q^2})}{q}}}{1-a^{-p}b^{p+q}t^{\frac{p (2 p + 3 q - \sqrt{4 p^2 - 3 q^2})}{q}}}+
\frac{a^{-p} b^{-p-q}t^{\frac{p (  3 q+2 p - \sqrt{4 p^2 - 3 q^2})}{q}}}{1-a^{-p}b^{-p-q}t^{\frac{p (2 p + 3 q - \sqrt{4 p^2 - 3 q^2})}{q}}}\\
&&+\frac{a^{p}b^{p-q}t^{\frac{p ( 3 q-2 p + \sqrt{4 p^2
- 3 q^2})}{q}}}{1-a^{p}b^{p-q}t^{\frac{p ( 3 q-2 p + \sqrt{4 p^2 -
3 q^2})}{q}}} +\frac{a^{p}b^{q-p}t^{\frac{p ( 3 q-2 p + \sqrt{4 p^2
- 3 q^2})}{q}}}{1-a^{p}b^{q-p}t^{\frac{p ( 3 q -2p+ \sqrt{4 p^2 - 3
q^2})}{q}}}\,. \nonumber \ee 
This is the left-handed index. The right-handed index is obtained by letting $a \to 1/a$.

\section{$T^{1,1}$ Index from Supergravity}\label{gravitysec}

On the dual supergravity side, the index of the conifold theory was 
computed by  Nakayama~\cite{Nakayama:2006ur}, using the results of \cite{Gubser:1998vd, Ceresole:1999zs, Ceresole:1999ht}
for the  KK reduction of IIB supergravity on $AdS_5 \times T^{1,1}$.

\begin{table}[!h]
\begin{centering}
\begin{tabular}{|l|l|l|l|l|l||l|}
\hline
\multicolumn{1}{|c|}{Fields} & Shortening Cond. & $s$ & $l$ & Mult. & $\II^{{\tt L}}(t,y)$ & \multicolumn{1}{l|}{$\sum_{\tilde{r}}({\cal I}_{[\tilde{r},j_2]_{+}^{{\tt L}}}^{{\tt L}}\times\ldots)$}\tabularnewline
\hline
\hline
Graviton & $E=1+\sqrt{H_{0}+4}$ & $\frac{r}{2}$ & $\frac{r}{2}$ & $\CC_{r(\frac{1}{2},\frac{1}{2})}$ & $\II_{[r+1,\frac{1}{2}]_{-}^{{\tt L}}}^{{\tt L}}$ & $-\chi_{\frac{\tilde{r}-1}{2}}(a)\chi_{\frac{\tilde{r}-1}{2}}(b)$\tabularnewline
\hline
\hline
Gravitino$_{{\rm I}}$  & $E=-\frac{1}{2}+\sqrt{H_{0}^{-}+4}$ & $\frac{r-1}{2}$ & $\frac{r-1}{2}$ & $\BB_{r(0,\frac{1}{2})}$ & $\II_{[r-2,\frac{1}{2}]_{-}^{{\tt L}}}^{{\tt L}}$ & $-\chi_{\frac{\tilde{r}+1}{2}}(a)\chi_{\frac{\tilde{r}+1}{2}}(b)$\tabularnewline
\cline{3-7}
 &  & $\frac{r-1}{2}$ & $\frac{r+1}{2}$ & $\CC_{r(0,\frac{1}{2})}$ & $\II_{[r,\frac{1}{2}]_{+}^{{\tt L}}}^{{\tt L}}$ & $+\chi_{\frac{\tilde{r}-1}{2}}(a)\chi_{\frac{\tilde{r}+1}{2}}(b)$\tabularnewline
\cline{3-7}
 &  & $\frac{r+1}{2}$ & $\frac{r-1}{2}$ & $\CC_{r(0,\frac{1}{2})}$ & $\II_{[r,\frac{1}{2}]_{+}^{{\tt L}}}^{{\tt L}}$ & $+\chi_{\frac{\tilde{r}+1}{2}}(a)\chi_{\frac{\tilde{r}-1}{2}}(b)$\tabularnewline
\hline
\hline
Gravitino$_{{\rm III}}$ & $E=-\frac{1}{2}+\sqrt{H_{0}^{+}+4}$ & $\frac{r+1}{2}$ & $\frac{r+1}{2}$ & ${\cal C}_{r(\frac{1}{2},0)}$ & ${\cal I}_{[r+1,0]_{-}^{{\tt L}}}^{{\tt L}}$ & $-\chi_{\frac{\tilde{r}}{2}}(a)\chi_{\frac{\tilde{r}}{2}}(b)$\tabularnewline
\hline
\hline
Gravitino$_{{\rm IV}}$ & $E=\frac{5}{2}+\sqrt{H_{0}^{-}+4}$ & $\frac{r-1}{2}$ & $\frac{r-1}{2}$ & ${\cal C}_{r(\frac{1}{2},0)}$ & ${\cal I}_{[r+1,0]_{-}^{{\tt L}}}^{{\tt L}}$ & $-\chi_{\frac{\tilde{r}}{2}-1}(a)\chi_{\frac{\tilde{r}}{2}-1}(b)$\tabularnewline
\hline
\hline
Vector$_{{\rm I}}$ & $E=-2+\sqrt{H_{0}+4}$ & $\frac{r}{2}$ & $\frac{r}{2}$ & ${\cal B}_{r(0,0)}$ & ${\cal I}_{[r-2,0]_{-}^{{\tt L}}}^{{\tt L}}$ & $-\chi_{\frac{\tilde{r}}{2}+1}(a)\chi_{\frac{\tilde{r}}{2}+1}(b)$\tabularnewline
\cline{3-7}
 &  & $\frac{r}{2}$ & $\frac{r+2}{2}$ & $\CC_{r(0,0)}$ & ${\cal I}_{[r,0]_{+}^{{\tt L}}}^{{\tt L}}$ & $+\chi_{\frac{\tilde{r}}{2}}(a)\chi_{\frac{\tilde{r}}{2}+1}(b)$\tabularnewline
\cline{3-7}
 &  & $\frac{r+2}{2}$ & $\frac{r}{2}$ & $\CC_{r(0,0)}$ & ${\cal I}_{[r,0]_{+}^{{\tt L}}}^{{\tt L}}$ & $+\chi_{\frac{\tilde{r}}{2}+1}(a)\chi_{\frac{\tilde{r}}{2}}(b)$\tabularnewline
\hline
\hline
Vector$_{{\rm IV}}$ & $E=1+\sqrt{H_{0}^{--}+4}$ & $\frac{r-2}{2}$ & $\frac{r-2}{2}$ & $\BB_{r(0,0)}$ & ${\cal I}_{[r-2,0]_{-}^{{\tt L}}}^{{\tt L}}$ & $-\chi_{\frac{\tilde{r}}{2}}(a)\chi_{\frac{\tilde{r}}{2}}(b)$\tabularnewline
\cline{3-7}
 &  & $\frac{r-2}{2}$ & $\frac{r}{2}$ & $\CC_{r(0,0)}$ & ${\cal I}_{[r,0]_{+}^{{\tt L}}}^{{\tt L}}$ & $+\chi_{\frac{\tilde{r}}{2}-1}(a)\chi_{\frac{\tilde{r}}{2}}(b)$\tabularnewline
\cline{3-7}
 &  & $\frac{r}{2}$ & $\frac{r-2}{2}$ & $\CC_{r(0,0)}$ & ${\cal I}_{[r,0]_{+}^{{\tt L}}}^{{\tt L}}$ & $+\chi_{\frac{\tilde{r}}{2}}(a)\chi_{\frac{\tilde{r}}{2}-1}(b)$\tabularnewline
\hline
\end{tabular}
\par\end{centering}
\caption{\label{gravity-multiplets}Short multiplets appearing in the KK reduction
of Type IIB supergravity on $AdS_{5}\times T^{1,1}$. In the last
column, we summarize the full index contributions of multiplets by
listing the $SU(2)_{s}\times SU(2)_{l}$ characters multiplying ${\cal I}_{[\tilde{r},\frac{1}{2}]_{+}^{{\tt L}}}^{{\tt L}}$
for first four rows and ${\cal I}_{[\tilde{r},0]_{+}^{{\tt L}}}^{{\tt L}}$
for remaining rows. The range of $\tilde{r}$ is specified by the two conditions that $\tilde{r}\geq -1$ {\it and} that the $SU(2)_{s}\times SU(2)_{l}$
representation makes sense. The chemical potentials $a$ and $b$
couple to $SU(2)_{s}\times SU(2)_{l}$ flavor charges respectively.
Exception: The first row of Gravitino$_{\rm{I}}$ starts from $\tilde{r}=0$. The $\tilde r =-1$ state of  Gravitino$_{\rm{I}}$
 gives rise to the Dirac multiplet ${\cal D}_{(0,\frac{1}{2})}$ due to additional shortening. It corresponds in the dual field theory
 to a decoupled $U(1)$ vector multiplet.
}
\end{table}

 Let us briefly review the structure of the calculation. For a general $AdS_5 \times Y^{p,q}$ background,
the KK spectrum organizes itself in three types of multiplets \cite{Ceresole:1999zs, Ceresole:1999ht}: graviton ($(\half,\half)$),
LH-gravitino ($(\half,0)$), RH-gravitino ($(0,\half)$), and vector ($(0,0)$).
 The details of the specific background manifest themselves
in the possible spectrum of the R-charges and their multiplicities. This information can be obtained by solving the spectrum of
relevant differential operators, {\it e.g.} scalar Laplacian and Dirac operators. For the $Y^{p,q}$ geometries the scalar
Laplacian is given by Heun's differential equation spectrum of which is hard to obtain in closed form,  see {\it e.g.}~\cite{Kihara:2005nt}.
For the $T^{1,1}$ background these data
were carefully computed in~\cite{Gubser:1998vd, Ceresole:1999zs, Ceresole:1999ht}. A generic multiplet of the KK spectrum
does not obey shortening conditions and thus does not contribute to the index.
 Table~\ref{gravity-multiplets} summarizes the multiplets which do contribute
 of the index.  The eigenvalue of the scalar laplacian is denoted by $H_0(s,l,r)$,
\be
H_0(s,l,r)=6(s(s+1)+l(l+1)-\frac{r^2}{8}).
\ee
$H_0^\pm$ and $H_0^{\pm\pm}$  are shorthands for $H_0(s,l,r\pm1)$ and $H_0(s,l,r\pm2)$ respectively.
Besides the KK modes of table~\ref{gravity-multiplets}, there are additional  Betti multiplets, arising
from the non-trivial homology of $T^{1,1}$. Their contribution to the index is found to vanish  \cite{Nakayama:2006ur}.

The $T^{1,1}$ manifold has $SU(2)_s\times SU(2)_l$ isometry. We refine the index by adding chemical potentials $a$ and $b$ that couple  respectively
to $SU(2)_s$ and $SU(2)_l$.
Simply reading off the R-charges and the multiplicities of the different modes,
we can write down the index as \cite{Nakayama:2006ur}\footnote{On the field theory side, we subtracted both $U(1)$ factors. Correspondingly, on the gravity side we should subtract all singleton degrees of freedom, and thus omit the $\tilde r = -1$ mode of the Gravitino$_{\rm{I}}$ tower, which corresponds
to a ${\cal D}_{(0,1/2)}$ multiplet.}
\begin{eqnarray}\label{gravityindex}
{\cal I}^{{\tt L}} & = &  -\sum_{\tilde{r}\geq0}{\cal I}_{[\tilde{r},\frac{1}{2}]_{+}^{{\tt L}}}^{{\tt L}}[(ab)^{\tilde{r}+1}+(\frac{a}{b})^{\tilde{r}+1}+(\frac{b}{a})^{\tilde{r}+1}+(\frac{1}{ab})^{\tilde{r}+1}] \nonumber\\
 & - & \sum_{\tilde{r}\geq-1}{\cal I}_{[\tilde{r},0]_{+}^{{\tt L}}}^{{\tt L}}[(ab)^{\tilde{r}}+(\frac{a}{b})^{\tilde{r}}+(\frac{b}{a})^{\tilde{r}}+(\frac{1}{ab})^{\tilde{r}}+(ab)^{\tilde{r}+2}+(\frac{a}{b})^{\tilde{r}+2}+(\frac{b}{a})^{\tilde{r}+2}+(\frac{1}{ab})^{\tilde{r}+2}] \nonumber\\
 & + & {\cal I}_{[-1,0]_{+}^{{\tt L}}}^{{\tt L}}\chi_{-\frac{3}{2}}(a)\chi_{-\frac{3}{2}}(b)-{\cal I}_{[0,0]_{+}^{{\tt L}}}^{{\tt L}}[-\chi_{-1}(a)\chi_{-1}(b)+\chi_{-1}(a)\chi_{0}(b)+\chi_{0}(a)\chi_{-1}(b)] 
 \end{eqnarray}
 The definition of the index  building blocks  ${\cal I}^{\tt L}_{[\tilde r, j_2]^{\tt L}_\pm}$ is given in the appendix,
 while the symbol $\chi_j(x)$ stands for the standard character of the spin-$j$ representation of $SU(2)$,
\be
\chi_{j}(x)\equiv \frac{x^{2j+1}-x^{-(2j+1)} }{x-x^{-1}} \,.
\ee
After simplification,
\be
{\cal I}^{{\tt L}}  =   \frac{t^{3}ab}{1-t^{3}ab}+\frac{t^{3}
\frac{a}{b}}{1-t^{3}\frac{a}{b}}+\frac{t^{3}\frac{b}{a}}{1-t^{3}
\frac{b}{a}}+\frac{t^{3}\frac{1}{ab}}{1-t^{3}\frac{1}{ab}}\, ,
\ee
which precisely agrees with  the large $N$ index (\ref{indT11}) computed from gauge theory using R\"omelsberger's prescription.

\section*{Acknowledgements}
It is a pleasure to acknowledge useful conversations with Davide Gaiotto, Ami Hanany, Zohar Komargodski, Juan Maldacena, Yu Nakayama and Yuji Tachikawa. This work was  supported in part by DOE grant DEFG-0292-ER40697 and by NSF grant PHY-0653351-001. Any
opinions, findings, and  conclusions or recommendations expressed in this
material are those of the authors and do not necessarily reflect the views of the National
Science Foundation.

\newpage

\appendix

\section{$\NN=1$ superconformal shortening conditions and the index}

In this appendix we summarize some basic facts about $\NN=1$
superconformal representation theory. 
 A generic long multiplet
$\AA^{\Delta}_{r(j_1, j_2)}$ is generated by the action of $4$
Poincar\'e supercharges $\QQ_\alpha$ and ${\widetilde \QQ}_{\dot \alpha}$
on superconformal primary which is by definition is annihilated by
all conformal supercharges $\cal S$.
In table \ref{N1-shortening} we have
summarized  possible shortening and semishortening conditions.

\begin{center}
\begin{table}[!h]
{\small
\begin{centering}
\begin{tabular}{|l|l|l|l|l|}
\hline
\multicolumn{4}{|c|}{Shortening Conditions} & Multiplet\tabularnewline
\hline
\hline
$\BB$ & $\QQ_{\alpha}|r\rangle^{h.w.}=0$ & $j_1=0$ & $\Delta=-\frac{3}{2}r$ & $\BB_{r(0,j_2)}$\tabularnewline
\hline
$\bar{\BB}$ & $\bar{\QQ}_{\dot{\alpha}}|r\rangle^{h.w.}=0$ & $j_2=0$ & $\Delta=\frac{3}{2}r$ & $\bar{\BB}_{r(j_1,0)}$\tabularnewline
\hline
$\hat{\BB}$ & $\BB\cap\bar{\BB}$ & $j_1,j_2,r=0$ & $\Delta=0$ & $\hat{\BB}$\tabularnewline
\hline
\hline
$\CC$ & $\e^{\alpha\beta}\QQ_{\beta}|r\rangle_{\alpha}^{h.w.}=0$ &  & $\Delta=2+2j_1-\frac{3}{2}r$ & $\CC_{r(j_1,j_2)}$\tabularnewline
 & $(\QQ)^{2}|r\rangle^{h.w.}=0$ for $j_1=0$ &  & $\Delta=2-\frac{3}{2}r$ & $\CC_{r(0,j_2)}$\tabularnewline
\hline
$\bar{\CC}$ & $\e^{\dot{\alpha}\dot{\beta}}\bar{\QQ}_{\dot{\beta}}|r\rangle_{\dot{\alpha}}^{h.w.}=0$ &  & $\Delta=2+2j_2+\frac{3}{2}r$ & $\bar{\CC}_{r(j_1,j_2)}$\tabularnewline
 & $(\bar{\QQ})^{2}|r\rangle^{h.w.}=0$ for $j_2=0$ &  & $\Delta=2+\frac{3}{2}r$ & $\bar{\CC}_{r(j_1,0)}$\tabularnewline
\hline
$\hat{\CC}$ & $\CC\cap\bar{\CC}$ & $\frac{3}{2}r=(j_1-j_2)$ & $\Delta=2+j_1+j_2$ & $\hat{\CC}_{(j_1,j_2)}$\tabularnewline
\hline
$\DD$ & $\BB\cap\bar{\CC}$ & $j_1=0,-\frac{3}{2}r=j_2+1$ & $\Delta=-\frac{3}{2}r=1+j_2$ & $\DD_{(0,j_2)}$\tabularnewline
\hline
$\bar{\DD}$ & $\bar{\BB}\cap\CC$ & $j_2=0,\frac{3}{2}r=j_1+1$ & $\Delta=\frac{3}{2}r=1+j_1$ & $\bar{\DD}_{(j_1,0)}$\tabularnewline
\hline
\end{tabular}
\par\end{centering}
} \caption{\label{N1-shortening}Possible shortening conditions 
for the $\NN=1$ superconformal algebra.}
\end{table}
\par\end{center}

A generic long multiplet of the ${\cal N}=1$ superconformal algebra
$SU(2,2|1)$ is $16(2j_1+1,2j_2+1)$ dimensional. 
Tables   \ref{N1Bmultiplets}, \ref{N1Cmultiplets}, \ref{hatC-multiplet} and \ref{D-multiplet} 
illustrate how the ${\cal B}$, ${\cal C}$, $\hat {\cal C}$ and ${\cal D}$-type  
multiplets fit within a generic  long multiplet. 

\newpage

\begin{table}[!h]
{\scriptsize
\begin{centering}
$\begin{array}{ccccccccccc}
\Delta &  &  &  &  &  & \boxed{(j_1,j_2)}\\
\\\Delta+\frac{1}{2} &  &  &  & (j_1+\frac{1}{2},j_2) &  &  &  & \boxed{(j_1,j_2+\frac{1}{2})}\\
 &  &  &  & (j_1-\frac{1}{2},j_2) &  &  &  & \boxed{(j_1,j_2-\frac{1}{2})}\\
 &  &  &  &  &  & (j_1+\frac{1}{2},j_2+\frac{1}{2})\\
\Delta+1 &  & \quad(j_1,j_2)\quad\qquad &  &  &  & (j_1-\frac{1}{2},j_2+\frac{1}{2}),(j_1+\frac{1}{2},j_2-\frac{1}{2}) &  &  &  & \qquad\quad\boxed{(j_1,j_2)}\\
 &  &  &  &  &  & (j_1-\frac{1}{2},j_2-\frac{1}{2})\\
\Delta+\frac{3}{2} &  &  &  & (j_1,j_2+\frac{1}{2}) &  &  &  & (j_1+\frac{1}{2},j_2)\\
 &  &  &  & (j_1,j_2-\frac{1}{2}) &  &  &  & (j_1-\frac{1}{2},j_2)\\
\\\Delta+2 &  &  &  &  &  & (j_1,j_2)\\
\\ &  & r-2 &  & r-1 &  & r &  & r+1 &  & r+2\end{array}$
\par\end{centering}}

\caption{\label{N1Bmultiplets}A long multiplet of $\NN=1$
superconformal algebra. The $SU(2,2)$ multiplets that are boxed form
a short $\BB_{r(0,j_2)}$ multiplet for $j_1=0, \Delta
=-\frac{3}{2}r$. The left-handed $\bar{\BB}$ can be obtained by
reflecting the table (that is, sending $r\to-r$ and $j_1\leftrightarrow
j_2$). In general, when $j_1(j_2)=0$, the $SU(2,2)$ multiplets
$(j_1-\frac{1}{2},any)((any,j_2-\frac{1}{2}))$ are set to zero,
resulting in further shortening.}

\end{table}

\begin{table}[!h]
{\scriptsize
\begin{centering}
$\begin{array}{ccccccccccc}
\Delta &  &  &  &  &  & \boxed{(j_1,j_2)}\\
\\\Delta+\frac{1}{2} &  &  &  & \boxed{(j_1+\frac{1}{2},j_2)} &  &  &  & \boxed{(j_1,j_2+\frac{1}{2})}\\
 &  &  &  & (j_1-\frac{1}{2},j_2) &  &  &  & \boxed{(j_1,j_2-\frac{1}{2})}\\
 &  &  &  &  &  & \boxed{(j_1+\frac{1}{2},j_2+\frac{1}{2})}\\
\Delta+1 &  & \quad(j_1,j_2)\quad\qquad &  &  &  & (j_1-\frac{1}{2},j_2+\frac{1}{2})\boxed{(j_1+\frac{1}{2},j_2-\frac{1}{2})} &  &  &  & \qquad\quad\boxed{(j_1,j_2)}\\
 &  &  &  &  &  & (j_1-\frac{1}{2},j_2-\frac{1}{2})\\
\Delta+\frac{3}{2} &  &  &  & (j_1,j_2+\frac{1}{2}) &  &  &  & \boxed{(j_1+\frac{1}{2},j_2)}\\
 &  &  &  & (j_1,j_2-\frac{1}{2}) &  &  &  & (j_1-\frac{1}{2},j_2)\\
\\\Delta+2 &  &  &  &  &  & (j_1,j_2)\\
\\ &  & r-2 &  & r-1 &  & r &  & r+1 &  & r+2\end{array}$
\par\end{centering}}

\caption{\label{N1Cmultiplets}
A long multiplet of $\NN=1$
superconformal algebra. The $SU(2,2)$ multiplets that are boxed form
a semi-short $\CC_{r(j_1,j_2)}$ multiplet for
$\Delta=2+2j_1-\frac{3}{2}r$. The left-handed $\bar{\CC}$ can be
obtained by reflecting the table (that is, sending $r\to-r$ and
$j_1\leftrightarrow j_2$). In general, when $j_1(j_2)=0$, the
$SU(2,2)$ multiplets $(j_1-\frac{1}{2},any)((any,j_2-\frac{1}{2}))$
are set to zero, resulting in further shortening.}

\end{table}
\begin{table}[!h]
{\scriptsize
\begin{centering}
$\begin{array}{ccccccccccc}
\Delta &  &  &  &  &  & \boxed{(j_1,j_2)}\\
\\\Delta+\frac{1}{2} &  &  &  & \boxed{(j_1+\frac{1}{2},j_2)} &  &  &  & \boxed{(j_1,j_2+\frac{1}{2})}\\
 &  &  &  & (j_1-\frac{1}{2},j_2) &  &  &  & (j_1,j_2-\frac{1}{2})\\
 &  &  &  &  &  & \boxed{(j_1+\frac{1}{2},j_2+\frac{1}{2})}\\
\Delta+1 &  & (j_1,j_2)\quad\qquad &  &  &  & (j_1-\frac{1}{2},j_2+\frac{1}{2})\quad(j_1+\frac{1}{2},j_2-\frac{1}{2}) &  &  &  & \qquad\quad(j_1,j_2)\\
 &  &  &  &  &  & \boxed{-(j_1-\frac{1}{2},j_2-\frac{1}{2})}\,\,\,\\
\Delta+\frac{3}{2} &  &  &  & (j_1,j_2+\frac{1}{2}) &  &  &  & (j_1+\frac{1}{2},j_2)\\
 &  &  &  & \boxed{-(j_1,j_2-\frac{1}{2})}\,\, &  &  &  & \boxed{-(j_1-\frac{1}{2},j_2)}\,\,\\
\\\Delta+2 &  &  &  &  &  & \boxed{-(j_1,j_2)}\,\,\\
\\ &  & r-2 &  & r-1 &  & r &  & r+1 &  & r+2\end{array}$
\par\end{centering}}

\caption{\label{hatC-multiplet}Multiplet structure of $\hat{\CC}_{(j_1,j_2)}$.
The shortening conditions are $\Delta=2+j_1+j_2$ and $\frac{3}{2}r=(j_1-j_2)$.}

\end{table}

\begin{table}[!h]
{\scriptsize
\begin{centering}
$\begin{array}{ccccccccccc}
\Delta &  &  &  &  &  & \boxed{(j_1,j_2)}\\
\\\Delta+\frac{1}{2} &  &  &  & (j_1+\frac{1}{2},j_2) &  &  &  & \boxed{(j_1,j_2+\frac{1}{2})}\\
 &  &  &  & (j_1-\frac{1}{2},j_2) &  &  &  & (j_1,j_2-\frac{1}{2})\\
 &  &  &  &  &  & (j_1+\frac{1}{2},j_2+\frac{1}{2})\\
\Delta+1 &  & (j_1,j_2)\quad\qquad &  &  &  & (j_1-\frac{1}{2},j_2+\frac{1}{2}),\boxed{-(j_1+\frac{1}{2},j_2-\frac{1}{2})} &  &  &  & \qquad\quad(j_1,j_2)\\
 &  &  &  &  &  & (j_1-\frac{1}{2},j_2-\frac{1}{2})\,\,\,\\
\Delta+\frac{3}{2} &  &  &  & (j_1,j_2+\frac{1}{2}) &  &  &  & \boxed{-(j_1+\frac{1}{2},j_2)}\\
 &  &  &  & (j_1,j_2-\frac{1}{2})\,\, &  &  &  & (j_1-\frac{1}{2},j_2)\,\,\\
\\\Delta+2 &  &  &  &  &  & (j_1,j_2),\boxed{+(j_1,j_2-1)}\\
\\\\\Delta+\frac{5}{2} &  &  &  &  &  &  &  & \boxed{+(j_1,j_2-\frac{1}{2})}\\
\\ &  & r-2 &  & r-1 &  & r &  & r+1 &  & r+2\end{array}$
\par\end{centering}}
\caption{\label{D-multiplet}Multiplet structure of $\DD_{(0,j_2)}$. The
shortening conditions are $\Delta=1+j_2=-\frac{3}{2}r$ and $j_1=0$.
The multiplet $\bar{\DD}_{(j_1,0)}$ could be obtained by $j_1 \leftrightarrow j_2,r\leftrightarrow-r$
or by simply reflecting the table. The shortening conditions in that
case are $\Delta=1+j_1=\frac{3}{2}r$ and $j_2=0$. }
\end{table}

%\pagebreak

At the unitarity threshold, a long multiplet
can decompose into  (semi)short multiplets. 
The splitting rules  are:
\begin{eqnarray*}
\AA_{r(j_1,j_2)}^{2+2j_1-\frac{3}{2}r} & \simeq & \CC_{r(j_1,j_2)}\oplus\CC_{r-1(j_1-\frac{1}{2},j_2)}\\
\AA_{r(j_1,j_2)}^{2+2j_2+\frac{3}{2}r} & \simeq & \bar{\CC}_{r(j_1,j_2)}\oplus\bar{\CC}_{r+1(j_1,j_2-\frac{1}{2})}\\
\AA_{\frac{2}{3}(j_1-j_2)(j_1,j_2)}^{2+j_1+j_2} & \simeq & \hat{\CC}_{(j_1,j_2)}\oplus\CC_{\frac{2}{3}(j_1-j_2)-1,(j_1-\frac{1}{2},j_2)}\oplus\bar{\CC}_{\frac{2}{3}(j_1-j_2)+1,(j_1,j_2-\frac{1}{2})}\end{eqnarray*}
We are using a notation where the
 $\BB$ and $\bar{\BB}$
type multiplets are formally identified with  special cases
of $\CC$ and $\bar{\CC}$ multiplets, as follows
\begin{equation}
\CC_{r(-\frac{1}{2},j_2)}\simeq\BB_{r-1(0,j_2)}\qquad\bar{\CC}_{r(j_1,-\frac{1}{2})}\simeq\bar{\BB}_{r+1(j_1,0)}\,.
\end{equation}
We define the ${\tt L}$eft (${\tt R}$ight) equivalence
class of the multiplet $\CC_{r(j_1,j_2)}(\bar{\CC}_{r(j_1,j_2)})$
as the class of multiplets with the same ${\tt L}$eft (${\tt R}$ight) index. From the splitting
rules, we see that the classes can be labeled as
 $[-r+2j_1,j_2]_{(-)^{2j_1}}^{{\tt L}}$ $([r+2j_2,j_1]_{(-)^{2j_2}}^{{\tt R}})$.
Moreover, ${\cal I}_{[-r+2j_1,j_2]_{-}^{{\tt L}}}^{{\tt L}}=-{\cal I}_{[-r+2j_1,j_2]_{+}^{{\tt L}}}^{{\tt L}}$
and ${\cal I}_{[r+2j_2,j_1]_{-}^{{\tt R}}}^{{\tt R}}=-{\cal I}_{[r+2j_2,j_1]_{+}^{{\tt R}}}^{{\tt R}}$.
The expressions for the indices of the  equivalent classes are
\begin{eqnarray*}
{\cal I}_{[\tilde{r},j_2]_{\pm}^{{\tt L}}}^{{\tt L}} & = & \pm(-)^{2j_2+1}\frac{t^{3(\tilde{r}+2)}\chi_{j_2}(y)}{(1-t^{3}y)(1-t^{3}y^{-1})}\\
{\cal I}_{[\bar{\tilde{r}},j_1]_{\pm}^{{\tt R}}}^{{\tt R}} & = & \pm(-)^{2j_1+1}\frac{t^{3(\bar{\tilde{r}}+2)}\chi_{j_1}(y)}{(1-t^{3}y)(1-t^{3}y^{-1})}\, \\
\II^{R} [\tilde{r},j_2]_{\pm}^{{\tt L}} &  =&  0 \\
 \II^{{\tt L}}[\bar{\tilde{r}},j_1]_{\pm}^{{\tt R}} & = & 0\,.
\end{eqnarray*}

The situation is slightly more involved for the $\hat{\CC}$ and $\DD$ type
multiplets. Unlike the $\BB,\CC$ type multiplets, they contribute both
to $\II^{{\tt L}}$ as well as $\II^{{\tt R}}$. The indices  \cite{Dolan:2008qi} for the different
types of multiplets are collected in table \ref{tab:Indices}.

\newpage

\begin{table}[htbp]
\begin{centering}
\begin{tabular}{|l|l|l|}
\hline
Multiplet & $\II^{{\tt L}}$ & $\II^{{\tt R}}$\tabularnewline
\hline
\hline
$\AA_{r(j_1,j_2)}^{\Delta}$ & $0$ & $0$\tabularnewline
\hline
$\CC_{r(j_1,j_2)}$ & $\II_{[-r+2j_1,j_2]_{(-)^{2j_1}}^{{\tt L}}}^{{\tt L}}$ & $0$\tabularnewline
\hline
$\bar{\CC}_{r(j_1,j_2)}$ & $0$ & $\II_{[r+2j_2,j_1]_{(-)^{2j_2}}^{{\tt R}}}^{{\tt R}}$\tabularnewline
\hline
\hline
$\hat{\CC}_{(j_1,j_2)}$ & $\II_{[\frac{2}{3}j_2+\frac{4}{3}j_1,j_2]_{(-)^{2j_1}}^{{\tt L}}}^{{\tt L}}$ & $\II_{[\frac{2}{3}j_1+\frac{4}{3}j_2,j_1]_{(-)^{2j_2}}^{{\tt R}}}^{{\tt R}}$\tabularnewline
\hline
$\DD_{(0,j_2)}$ & $\II_{[\frac{2}{3}j_2-\frac{4}{3},j_2]_{-}^{{\tt L}}}^{{\tt L}}+\II_{[\frac{2}{3}j_2-\frac{1}{3},j_2-\frac{1}{2}]_{-}^{{\tt L}}}^{{\tt L}}$ & $\II_{[\frac{4}{3}j_2-\frac{2}{3},0]_{+}^{{\tt R}}}^{{\tt R}}$\tabularnewline
\hline
$\bar{\DD}_{(j_1,0)}$ & $\II_{[\frac{4}{3}j_1-\frac{2}{3},0]_{+}^{{\tt L}}}^{{\tt L}}$ & $\II_{[\frac{2}{3}j_1-\frac{4}{3},j_1]_{-}^{{\tt R}}}^{{\tt R}}+\II_{[\frac{2}{3}j_1-\frac{1}{3},j_1-\frac{1}{2}]_{-}^{{\tt R}}}^{{\tt R}}$\tabularnewline
\hline
\end{tabular}
\par\end{centering}

\caption{\label{tab:Indices}Indices $\II^{{\tt L}}$ and $\II^{{\tt R}}$ of the various
short and semi-short multiplets.}

\end{table}

%\newpage

\bibliography{sduality}
%\bibliography{ref}
\bibliographystyle{JHEP}
%\bibliographystyle{unsrt}
%\bibliography{ref}
%\bibliographystyle{apsrev}
%\bibliographystyle{plain}
%\bibliographystyle{utphys}

\end{document}